\documentclass[aps,prc,twocolumn,superscriptaddress,nofootinbib]{revtex4}
\usepackage{graphicx}
\usepackage{amsfonts,amsmath,color,epsfig,epstopdf}
\usepackage{float}
\usepackage{multirow}
\usepackage[normalem]{ulem}
\usepackage{tabularx}
\usepackage{bbm} %blackboard bold numerals
\usepackage[caption=false,justification=justified,singlelinecheck=false]{subfig}

\newcommand{\SU}[1]{\ensuremath{\mathrm{SU}( #1 )}}
\newcommand{\Un}[1]{\ensuremath{\mathrm{U}( #1 )}}

\newcommand{\SpR}[1]{\ensuremath{\mathrm{Sp}( #1,\mathbb{R} )}}
\newcommand{\abinitio}{\emph{ab initio }}
\newcommand{\half}{\ensuremath{\textstyle{\frac{1}{2}}}} % 1/2 fraction small size
\newcommand\nuc[2]{\ensuremath{\mathrm{\textsuperscript{#1}#2}}} % nuclear notation

\newcommand{\ho}{\ensuremath{\hbar\Omega}} % hbarOmega symbol
\newcommand{\gs}{ground state} % ground state gs is italicized
\newcommand{\Nmax}{\ensuremath{N_{\mathrm{max}}}} % shorthand Nmax notation
\begin{document}

\title{Alpha clustering and alpha-capture reaction rate \\
from  \textit{ab initio}  symmetry-adapted  description of \nuc{20}{Ne}
 }
\author{A. C. Dreyfuss}
\affiliation{Department of Physics and Astronomy, Louisiana State University, Baton Rouge, LA 70803, USA}

\author{K. D. Launey}
\affiliation{Department of Physics and Astronomy, Louisiana State University, Baton Rouge, LA 70803, USA}

\author{J. E. Escher}
\affiliation{Lawrence Livermore National Laboratory L-414, Livermore, California 94551, USA}

\author{G. H. Sargsyan}
\affiliation{Department of Physics and Astronomy, Louisiana State University, Baton Rouge, LA 70803, USA}

\author{R. B. Baker}
\affiliation{Department of Physics and Astronomy, Louisiana State University, Baton Rouge, LA 70803, USA}

\author{T. Dytrych}
\affiliation{Department of Physics and Astronomy, Louisiana State University, Baton Rouge, LA 70803, USA}
\affiliation{Nuclear Physics Institute, 250 68 $\check{R}$e$\check{z}$, Czech Republic}

\author{J. P. Draayer}
\affiliation{Department of Physics and Astronomy, Louisiana State University, Baton Rouge, LA 70803, USA}

\begin{abstract}
We introduce a new framework for studying clustering and  for calculating alpha partial widths using \abinitio wave functions. We demonstrate the formalism for \nuc{20}{Ne}, by calculating  the overlap between the $\nuc{16}{O}+\alpha$ cluster configuration and states in \nuc{20}{Ne} computed  in the $\abinitio$ symmetry-adapted no-core shell model. We present spectroscopic amplitudes and spectroscopic factors, and compare those to  no-core symplectic shell-model results in larger model spaces, to gain insight into the underlying physics that drives alpha-clustering.   Specifically, we report on the alpha partial width of the lowest $1^-$ resonance in \nuc{20}{Ne}, which is found to be in good agreement with experiment. We also present first  no-core shell-model estimates for asymptotic normalization coefficients for the ground state, as well as for the first excited  $4^{+}$ state in \nuc{20}{Ne} that lies in a close proximity to the $\alpha + \nuc{16}{O}$ threshold.
This outcome highlights the importance of correlations for developing cluster structures and for describing alpha widths.
The widths can  then  be used to calculate alpha-capture reaction rates for narrow resonances of interest to astrophysics.  We explore the  reaction rate for the alpha-capture reaction $\nuc{16}{O}(\alpha,\gamma)\nuc{20}{Ne}$ at astrophysically relevant temperatures and determine its impact on simulated X-ray burst abundances.  
\end{abstract}

\maketitle

%%%%%%%%%%%%%%%%%%%%
%%% Introduction %%%
%%%%%%%%%%%%%%%%%%%%
\section{Introduction}
Modeling  nuclear systems with cluster substructure represents a major challenge for many-particle approaches that build on realistic interactions, such as those derived in the chiral effective field theory ($\chi$EFT) framework \cite{Machleidt_PR503_2011,Epelbaum_PPNP57_2006}. The earliest techniques for describing clustering use an underlying assumption of clusters in a few-body framework with microscopic interactions. For example, microscopic cluster models (MCMs), such as the resonating group method (RGM) \cite{Wheeler_PR52_1937a, Wheeler_PR52_1937b} and the related generator coordinate method (GCM) \cite{Horiuchi_PTP43_1970}, treat all particles within localized clusters. These MCM approaches have been used to study various reactions of astrophysical importance, including the first studies of $\alpha$-$\alpha$ scattering within the RGM framework \cite{Tang_PR47_1978}. 
Studies of the Hoyle state in \nuc{12}{C} and alpha conjugate nuclei (that is, nuclei with multiples of 2 protons and 2 neutrons)  have been of a special interest. E.g., the  antisymmetrized molecular dynamics (AMD) \cite{Enyo_PRL81_1998} and fermionic molecular dynamics (FMD) \cite{Chernykh_PRL98_2007} methods have been used to study the \nuc{12}{C} Hoyle state and its rotational band, and both saw evidence of these states as an extended 3-$\alpha$ cluster system. The THSR model \cite{Funaki_PPNP82_2015} describes alpha conjugate nuclei as condensates of $\alpha$ particles, and has been used to describe states in \nuc{12}{C}, \nuc{16}{O}, and of \nuc{20}{Ne} (for a review of cluster models, see Ref. \cite{Freer_RMP90_2018}). 

The microscopic cluster basis used in the RGM has a complementary nature  \cite{Hecht_PRC16_1977,Hecht_NPA318_1979} to the \SpR{3} symplectic basis, with \SpR{3} the underpinning symmetry of the microscopic symplectic model \cite{Rosensteel_PRL38_1977,Rowe_RPP48_1985}, the no-core symplectic shell model (NCSpM) \cite{Dreyfuss_PLB727_2013,Tobin_PRC89_2014,Dreyfuss_PRC95_2017}, and the symmetry-adapted no-core shell model (SA-NCSM) \cite{ Launey_PPNP89_2016,DytrychLDRWRBB20}. A number of studies have taken advantage of that relationship using a single \SU{3} irreducible representation (irrep) for the clusters\footnote{The deformation-related \SU{3} group is a subgroup of the symplectic \SpR{3} group, which preserves an equilibrium shape and its rotations and vibrations \cite{Rowe_AIP1541_2013,DytrychLDRWRBB20}.}. In particular, this approach has been used to describe the sub-Coulomb $\nuc{12}{C}+\nuc{12}{C}$ resonances of \nuc{24}{Mg} \cite{Suzuki_NPA388_1982} of particular interest in astrophysics \cite{Wiescher_book}, and overlaps between symplectic and cluster states for alpha conjugate nuclei \cite{Suzuki_NPA448_1986,Suzuki_NPA455_1986} which have been used to compute spectroscopic amplitudes \cite{Hecht_NPA244_1975,Hecht_NPA356_1981,Suzuki_KDLbook}.
These studies have shown that some of the most important shell-model configurations can be expressed by exciting the relative-motion degree of freedom of the clusters. Further, they have indicated that an approach that utilizes both the cluster and symplectic bases proves to be advantageous \cite{Hecht_PLB103_1981}, especially since the model based on the cluster basis only, for clusters without excitations, tends to overestimate cluster decay widths and underestimates $E2$ transition rates \cite{Suzuki_book2003}.

In this paper, we outline a new many-body technique for determining challenging alpha widths and asymptotic normalization coefficients (ANCs), with applications to \nuc{20}{Ne}, by  using \abinitio SA-NCSM  wave functions.  The formalism  builds on the complementary nature of the symplectic basis and the cluster basis. The SA-NCSM is ideal for addressing cluster substructures,  as it enables the reach of intermediate-mass nuclei and large model spaces by exploiting the symmetry-adapted basis \cite{DytrychLDRWRBB20}. We compare the outcome to results from the NCSpM with an effective many-nucleon interaction, which can reach ultra-large model spaces, and has achieved successful no-core shell-model descriptions of low-lying states in deformed $A=8-24$ nuclei \cite{Tobin_PRC89_2014}, and in particular, the  Hoyle state in $^{12}$C and its first $2^+$ and $4^+$ excitations \cite{Dreyfuss_PLB727_2013,Launey_PPNP89_2016,Dreyfuss_PRC95_2017}. 
There  has been recent progress in $\abinitio$ descriptions of alpha cluster systems, e.g., Green's Function Monte Carlo (GFMC) method with applications to the $\alpha$-cluster structure of \nuc{8}{Be} and \nuc{12}{C}, along with electromagnetic (EM) transitions \cite{Wiringa_PRC62_2000,Carlson_RMP87_2015}; the nuclear lattice effective field theory (NLEFT) with applications to the Hoyle state energy and the astrophysically relevant $\alpha$-$\alpha$ scattering problem \cite{Epelbaum_PRL106_2011,Rupak_PRL111_2013,Elhatisari_Nature528_2015}; and the hyperspherical harmonics (HH) method, with applications to giant resonance modes in \nuc{4}{He} \cite{Bacca_PRL110_2013}. Of particular note are recent developments that combine RGM with configuration-interaction methods \cite{Volya_PRL119_2017,Mercenne_PRC99_2019}, as well as with $\abinitio$ no-core shell model and SA-NCSM \cite{Quaglioni_PRL101_2008,Navratil_JPGNP36_2009,Barrett_PPNP69_2013,Mercenne_CNR}.

Specifically, we provide first results of the alpha partial width of the first excited $1^-$ state in the well-studied, highly-clustered \nuc{20}{Ne} system, starting from \abinitio calculations of the $A$-body system. While the present framework is general, as a first step, we assume here that each of the two clusters, $\alpha$ and $\nuc{16}{O}$, has a single equilibrium shape with suppressed vibrations.  We focus on the 5.79-MeV
 $1^{-}$ excited  state  in \nuc{20}{Ne}, which dominates the alpha-capture reaction rate for the $\nuc{16}{O}(\alpha,\gamma)\nuc{20}{Ne}$ reaction at astrophysical temperatures.  The natural width of the $1^{-}$ resonance is known 
 \cite{PhysRevC.22.356}, 
 and since the state decays entirely through $\alpha$ emission, the natural width is the $\alpha$ partial width. We note that partial widths are not directly measurable, and extraction is (to larger or smaller extent) model-dependent. 
The $\alpha$-cluster structure (through the $\alpha+\nuc{16}{O}$ partitioning) of \nuc{20}{Ne} is very well-studied. From a theoretical perspective, particular attention has been paid to whether the low-lying positive- and negative-parity rotational bands in \nuc{20}{Ne} are a pair of inversion doublet rotational bands of $\alpha+\nuc{16}{O}$ \cite{Horiuchi_PTPS192_2012,Zhou_PRL110_2013}. Early experiments have determined spectroscopic factors through fitting data from transfer reactions, specifically, $\nuc{16}{O}(\nuc{6}{Li},\,d)\nuc{20}{Ne}$ and its inverse reaction $\nuc{20}{Ne}(d,\,\nuc{6}{Li})\nuc{16}{O}$, to distorted wave Born approximation (DWBA) calculations \cite{Becchetti_NPA305_1978,Anantaraman_NPA313_1979,Tanabe_PRC24_1981,Oelert_PRC459_1979}. A more recent study re-analyzed this data within a coupled-channel Born approximation framework to reduce ambiguity in the fitting procedure due to uncertainty in the \nuc{6}{Li} optical model potential \cite{Fukui_PRC93_2016}. Experimental methods are moving away from the use of spectroscopic factors to extract information about clustering and the $\alpha$ partial widths; some instead use a method that extracts alpha partial widths through the ANCs, a technique that was first formulated for and applied to single-particle projectiles \cite{Mukhamedzhanov_PRC56_1997,Tribble_RPP77_2014}. It has since been applied to both the $\alpha+\nuc{12}{C}$ cluster structure of \nuc{16}{O} \cite{Brune_PRL83_1999,Avila_PRL114_2015} as well as to the $\alpha+\nuc{16}{O}$ cluster structure of \nuc{20}{Ne} \cite{Avila_PRC90_2014}, to achieve exceptional agreement with the accepted value of the width.

The present outcomes indicate pronounced clustering associated with an excited $1^{-}$ state in \nuc{20}{Ne}, as well as 
less pronounced clusters
 in the ground state, in agreement with previous results \cite{Buck_PRC11_1975,KanadaEnyo_PTEP2014_2014}. The calculated $1^{-}$ alpha partial width,  with its uncertainties, is found to be in good agreement with the experiment \cite{Constantini_PRC82_2010}, given that no parameters  have been adjusted in this study.  The resulting deviation between the theoretical and experimental values is examined in  X-ray burst (XRB) simulations. For this, we use the  width to estimate the largest direct contribution to the $\nuc{16}{O}(\alpha,\gamma) \nuc{20}{Ne}$  reaction rate at the astrophysically relevant energy regime. Indeed, we find almost no difference in the XRB abundance pattern when one uses the experimental width or the calculated width.  
Hence, this method is important to estimate reaction rates for reactions of astrophysical significance that are difficult or impossible to measure. For example, a work in progress focuses on  the $\nuc{15}{O}(\alpha, \gamma)\nuc{19}{Ne}$ reaction, which has been suggested to have the highest impact on XRB nucleosynthesis \cite{Cyburt_ApJ830_2016}.

%%%%%%%%%%%%%%%%%%%%    
%%%%%% Theory %%%%%%
%%%%%%%%%%%%%%%%%%%%
\section{Many-body theoretical framework}
\label{subsec:coords}
\begin{figure}[t]  
    \includegraphics[width=0.27\textwidth]{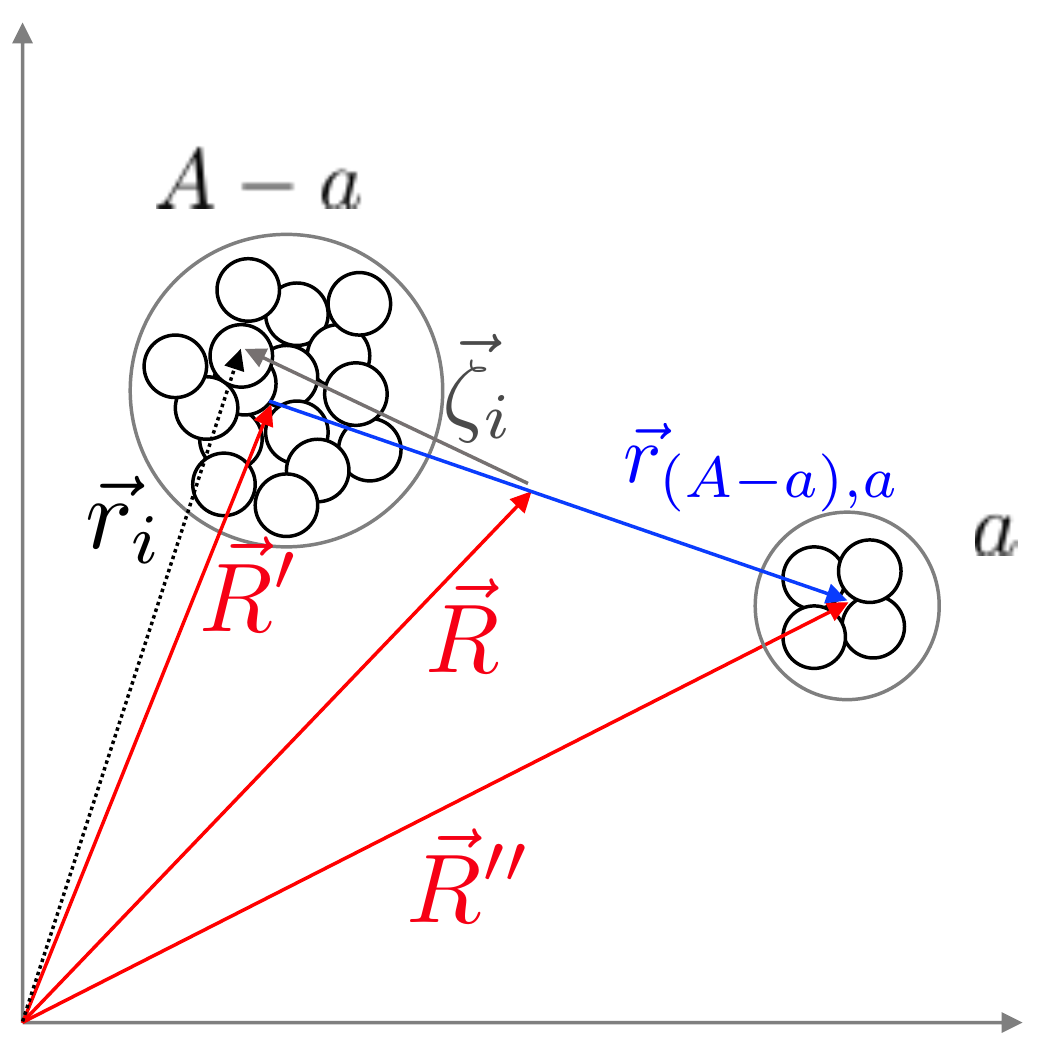}
    \caption
    {
        \label{fig:Coords}
        Illustration of the coordinates used in the formalism:  $\vec{r}_{i}$ are the  particle coordinates in the laboratory frame; $\vec{R'}$ and $\vec{R''}$ are the center-of-mass coordinates of the individual clusters, $\vec{R}$ is the center-of-mass coordinate of the $A$-body system;  $\vec{\zeta}_{i}$ are the relative coordinates with respect to $\vec{R}$. The relative separation of the clusters is $\vec{r}_{(A-a),a}$. 
    }
\end{figure}

For a system of $A$ particles, the set of laboratory coordinates is denoted as $\vec{r}_{1},\dots,\vec{r}_{A}$. A two-cluster system, with an $(A-a)$-particle cluster and an $a$-particle cluster separated by $\vec{r}_{A-a,a}$, can be divided into two distinct sets of laboratory coordinates, $\vec{r}_1,\dots,\vec{r}_{A-a}$ and $\vec{r}_{A-a+1},\dots,\vec{r}_{A}$ (see Fig. \ref{fig:Coords}). The centers of mass of the two clusters ($\vec{R'}$ and $\vec{R''}$) and the composite system $\vec{R}$, along with the distance between the clusters $\vec{r}_{A-a,a}$ are given as:
\begin{align}
    & \vec{R'} = \frac{1}{A-a} \sum_{i=1}^{A-a} \vec{r}_{i}, \,
     \vec{R''} = \frac{1}{a} \sum_{i=A-a+1}^{A} \vec{r}_{i} ,\\
    & \vec{R}  = \frac{1}{A} \sum_{i=1}^{A} \vec{r}_{i} = \frac{(A-a)\vec{R'}+a\vec{R''}} {A},  \label{EQ:CoMR}\\
    &\vec{r}_{A-a,a}  =\vec{R''}-\vec{R'}. \label{r}
\end{align}
The derivations in this work are based on relative coordinates with respect to the center-of-mass (CM) of the $A$-particle system,
\begin{equation}
    \label{EQ:relxi}
    \vec{\zeta}_{i}=\vec{r}_{i}-\vec{R}.
\end{equation}
The two clusters can be written in terms of relative coordinates $\vec{\boldsymbol{\zeta}'}\equiv\vec{\zeta}_{1},\dots,\vec{\zeta}_{A-a}$ and $\vec{\boldsymbol{\zeta}''}\equiv\vec{\zeta}_{A-a+1},\dots,\vec{\zeta}_{A}$, respectively. Note, from Eqs. (\ref{EQ:CoMR}) and (\ref{EQ:relxi}), that $\sum_{i=1}^{A} \vec{\zeta}_{i} = 0$, so that there are only $A-1$ independent relative coordinates. The translationally invariant two-cluster system is fully described by the relative coordinates 
\begin{equation}
    \vec{\boldsymbol{\zeta}}=\{ \vec{\zeta}_{1},\dots,\vec{\zeta}_{A-a},\vec{\zeta}_{A-a+1},\dots,\vec{\zeta}_{A-1}\}.
\end{equation}

The cluster system is defined for a channel $\nu$, which is given by the spin and parity of each of the clusters $\nu  = \{ \mathfrak{a}, \mathfrak{a}', I^{\pi'}, \mathfrak{a}'', I''^{\pi''}\}$ (the labels $\mathfrak{a}$, $\mathfrak{a}'$ and $\mathfrak{a}''$ denote all other quantum numbers needed to fully characterize their respective states), and a partial wave $l$, or the orbital momentum of the relative motion of the clusters, and has a good total angular momentum and parity, $J^{\pi}$, given by the coupling of the total angular momentum $I$ of the clusters to $l$.

\subsection{Resonances}
\label{sec:res}
The partial width of a resonance state corresponding to the emission of an $a$-particle cluster with relative angular momentum $l$ is given by \cite{Lovas_PR294_1998, Thomas_PTP12_1954}
\begin{equation}
    \label{eq:width}
    \Gamma_{a}(r_{c}) = 2P_{l}(r_{c}) \frac{\hbar^2}
    {2\mu_{A-a,a} r_{c}} \left[ ru_{\nu Il}^{J^{\pi}}(r) \right]^{2}_{r=r_{c}},
\end{equation}
where 
\begin{equation}
\mu_{A-a,a} = {m_{A-a}m_{a}\over m_{A-a}+m_{a} }
\label{redmass}
\end{equation}
is the reduced mass of the system with $m_{a}$ being the mass of the $a$-particle cluster, and $m_{A-a}$ being the mass of the $(A-a)$-particle cluster. $P_{l}(r)$ is the Coulomb penetrability, 
 $ru_{\nu Il}^{J^{\pi}}(r)$ is the spectroscopic amplitude (sometimes called the ``formation amplitude'' \cite{Lovas_PR294_1998}), and  the separation distance  $r_{c}$ between the two clusters is the channel radius.  
Eq. (\ref{eq:width}) is based on a well-established $R$-matrix theory \cite{Descouvement_RPP73_2010}, where the  channel radius $r_{c}$ is the separation distance, at which the interior nuclear wave function and exterior, Coulomb-dominated, wave function are matched. The new feature in this study is a novel approach to calculating the spectroscopic amplitude $ru_{\nu Il}^{J^{\pi}}(r)$ within a many-body no-core shell-model framework (Sec. \ref{SpecAmp}).

The partial width can also be written through the reduced width,
\begin{equation}
  \label{eq:width2}
\gamma^{2}_{\nu l}(r_{c})=\frac{
\hbar^2
}{2\mu_{A-a,a} r_{c}}\left[ ru_{\nu Il}^{J^{\pi}}(r) \right]^{2}_{r=r_{c}},
\end{equation}
as $ \Gamma_{a}(r_{c}) = 2P_{l}(r_{c})\gamma^{2}_{\nu l}(r_{c})$, where the penetrability is driven by the Coulomb force at large distances, while the reduced width $\gamma^{2}_{\nu l}(r_{c})$ contains information about the wave function at small distances. Experimental studies often report a unitless reduced width, relative to the Wigner limit $\gamma^{2}_{W}( r_c )=3\hbar^{2}/2\mu r_c^{2}$ \cite{Teichmann_PR87_1952}:
\begin{equation}
    \label{eq:unitlessredwide}
    \theta_{\nu l}^{2} ( r_c )
    =
    \gamma^{2}_{\nu l}( r_c )
    /
    \gamma^{2}_{W}(  r_c ).
  \end{equation}
The norm of the spectroscopic amplitude,
\begin{equation}
    \label{eq:SF}
    SF=\int_{0}^{\infty}|ru_{\nu Il}^{J^{\pi}}(r)|^{2}dr,
\end{equation}
is called the spectroscopic factor. 

The Coulomb penetrability $P_{l}(r)$ is determined by $H_{l}^{+}(\eta_{R},kr)$, the outgoing spherical Hankel function solution to the Coulomb equation defined by the Sommerfeld parameter $\eta_{R}=\frac{Z_{A-a}Z_{a}\mu_{A-a,a} e^2}{\hbar^2 k}$ for two clusters of charge $Z_{A-a}$ and $Z_{a}$,
\begin{equation}
    P_{l}(r) = \frac{kr}{|H_{l}^{+}(\eta_{R},kr)|^{2}},
\end{equation}
where the momentum $k$ corresponds to the positive energy in the center-of-mass frame
$E=\frac{\hbar^2k^2}{2\mu_{A-a,a}}$. 

\subsection{Bound states}
For bound states, the exterior wave function is given by an asymptotically-decaying Whittaker function:
\begin{equation}
    W_{-\eta_{B},l+\frac{1}{2}}(2\kappa_{B}r) \,\overrightarrow{r\rightarrow\infty}\,(2\kappa_{B}r)^{-\eta_{B}}e^{-\kappa_{B}r}.
\end{equation}
Here, $\kappa_{B}=\sqrt{-2\mu_{A-a,a} E}/\hbar$, for a negative energy $E$, and $\eta_{B}=\frac{Z_{A-a}Z_{a}\mu_{A-a,a} e^2}{\hbar^2 \kappa_{B}}$ is the associated Sommerfeld parameter. The observable ANC ($C_{l}$) determines the amplitude of the exterior wave function at large distances $r$, so that the exterior bound state wave function is written as \cite{Nuclear_Reactions_textbook}
\begin{equation}
    \label{eq:ext}
    \phi_{\nu Il}^{J^{\pi},\,\mathrm{ext}}(r) = C^{J^{\pi}}_{\nu Il}W_{-\eta_{B},l+\frac{1}{2}}(2\kappa_{B}r).
\end{equation}
The ANC can be determined through matching the exterior wave function with the interior bound state wave function,
\begin{equation}
    \label{eq:int}
    \phi_{\nu Il}^{J^{\pi},\,\mathrm{int}}(r)= C_{\rm int} ru_{\nu Il}^{J^{\pi}}(r),
\end{equation}
where $C_{\rm int}$ is the norm of the interior wave function, and in $R$-matrix theory, $\int_0^{r_c }|ru_{\nu Il}^{J^{\pi}}(r)|^2 dr=1$. Since the complete (interior + exterior) wave function must be normalized to unity, the norm of the interior contribution is given as
\begin{equation}
    C_{\rm int}^{2}=1-(C^{J^{\pi}}_{\nu Il})^{2}\int^{\infty}_{r_{c}}|W_{-\eta_{B},l+\frac{1}{2}}(2\kappa_{B}r)|^{2}dr.
\end{equation}
Matching the interior and exterior solutions at the channel radius $r_{c}$ yields the following expression for the ANC:
\begin{align}
    \label{eq:ANC}
   (C^{J^{\pi}}_{\nu Il})^{-2}=  \left.  \frac{ \big\rvert W_{-\eta_{B},l+\frac{1}{2}}(2\kappa_{B}r)\big\rvert^{2}}{\big\rvert ru_{\nu Il}^{J^{\pi}}(r)\big\rvert^{2} } \right|_{r_c} \nonumber \\
    +\int^{\infty}_{r_{c}}|W_{-\eta_{B},l+\frac{1}{2}}(2\kappa_{B}r)|^{2}dr.
\end{align}
We emphasize that determining accurate ANCs with this equation often requires a large $r_c$ to ensure the accurate treatment of the interior wave function, which may be impractical for \abinitio or microscopic models. Alternative ways to calculate ANCs based on a method with a faster ANC convergence are discussed in Refs. \cite{Nollet_PRC83_2011,Brune_PRC66_2002}.

\subsection{Spectroscopic Amplitudes}
\label{SpecAmp}
As discussed above, determining either the resonance partial width (\ref{eq:width}) or the ANC of a bound state (\ref{eq:ANC}) requires a calculation of the spectroscopic amplitude. The spectroscopic amplitude is given through the overlap of the composite $A$-particle state $\Psi_{(A)}$ and the cluster state. For the partition into $a$- and $(A-a)$-particle clusters, with intrinsic wave functions $\psi_{(a)}$ and $\psi_{(A-a)}$, the spectroscopic amplitude $ru_{\nu Il}^{J^{\pi}}(r)$ is given by (see, e.g., \cite{Lovas_PR294_1998,Nollet_PRC83_2011})
\begin{equation}
    \label{eq:specamp1}
    u_{\nu Il}^{J^{\pi}}(r) = \sum_{M_{I}m} C_{IM_{I}lm}^{JM} \int d\hat{r} Y^{*}_{lm}(\hat{r})u_{\nu}^{IM_{I}}(\vec{r}),
\end{equation}
where $C_{IM_{I}lm}^{JM}$ are Clebsch-Gordan coefficients and
\begin{align}
    &u_{\nu}^{IM_{I}}(\vec{r})
     = \int d^{3}\vec{\boldsymbol\zeta}\,[ \Psi^{\mathfrak{a} J^{\pi} M }_{(A)} (\vec{\boldsymbol\zeta})]^{\dagger}
    \\
    & \times \mathcal{A}[\{ \psi^{\mathfrak{a}'I'^{\pi'}}_{(A-a)}(\vec{\boldsymbol\zeta'}) \times \psi^{\mathfrak{a}''I''^{\pi''}}_{(a)}(\vec{\boldsymbol\zeta''}) \}^{IM_{I}} \delta(\vec{r}-\vec{r}_{A-a,a})]\nonumber.
\end{align}
In the cluster wave function, the operator $\mathcal{A}$ ensures that particles are properly antisymmetrized among the two clusters.

The delta function depends on the relative distance between the clusters and can be expanded in HO wave functions (with HO frequency $\hbar\Omega$) for the relative motion of an effective single particle with a reduced mass  (\ref{redmass}), which implies the use of a HO constant $b_{\mathrm{rel}} = \sqrt{\hbar/\mu_{A-a,a}\Omega}$. The expansion is given in the HO radial functions $R_{\eta lm}(r)$ (with $\eta l m$ denoting the HO shell number, orbital momentum and its projection, respectively):
\begin{align}
    \delta&(\vec{r}-\vec{r}_{A-a,a})\\\nonumber
    &= \sum_{\eta l'm'} R_{\eta l'}(r) Y_{l'm'}(\hat{r})R_{\eta l'}(r_{A-a,a})Y_{l'm'}(\hat{r}_{A-a,a}).
\end{align}
Using this and inserting $\int\phi^{*}_{000}(\vec{R})\phi_{000}(\vec{R}) d^{3}R = 1$ for the center-of-mass wave function $\phi_{\eta_{\rm CM }=0\,l_{\rm CM}=0\, m_{\rm CM}=0}(\vec{R})$ that is in the lowest-energy HO configuration, Eq. (\ref{eq:specamp1}) becomes,
\begin{align}
    \label{eq:SAwithOverlap}
    &u_{\nu Il}^{J^{\pi}}(r)
    =
    \sum_{\eta } R_{\eta l}(r)
    \int d^{3}\vec{\boldsymbol{\zeta}} d^3R
    [ \Psi^{\mathfrak{a} J^{\pi} M}_{(A)} (\vec{\boldsymbol\zeta})\phi_{000}(\vec{R})]^{\dagger}
    \nonumber
    \\
    &
    \times
    \mathcal{A}[\{ \psi^{\mathfrak{a}'I'^{\pi'}}_{(A-a)}(\vec{\boldsymbol\zeta'}) \times \psi^{\mathfrak{a}''I''^{\pi''}}_{(a)}(\vec{\boldsymbol\zeta''}) \}^{I}
    \nonumber
    \\
    &
    ~~~~~~~~~~~~~~~~~~~~~~~~~~~~~~
    \times   
    \chi_{\eta l}(\vec{r}_{A-a,a})\phi_{000}(\vec{R})]^{J^{\pi}M} 
    \nonumber \\
        &=
    \sum_{\eta } R_{\eta l}(r)
    \langle
    (A)\mathfrak{a} J^{\pi}M |(\mathfrak{a}'I'^{\pi'},\mathfrak{a}''I''^{\pi''})I,\eta l;J^{\pi}M\rangle,
\end{align}
where 
$\chi_{\eta lm}(\vec{r}_{A-a,a})=R_{\eta l}(r_{A-a,a})Y_{lm}(\hat{r}_{A-a,a})$ 
is the relative motion of the two many-body clusters at a separation distance defined by the particle coordinates $\vec{\boldsymbol{\zeta}}$ (\ref{r}). The wave functions of the composite $A$-particle system and the cluster system are given
with an explicitly separable center-of-mass contribution $\phi_{000}(\vec{R})$. This is important since the  wave function of the $A$-particle system is solved here in the no-core shell model, where spurious center-of-mass excitations are removed through a Lawson procedure. The resulting final eigenfunctions are each exactly factorized to a CM contribution, which is in the lowest HO energy, and an intrinsic wave function.

Eq. (\ref{eq:SAwithOverlap}) provides a transition from the many-body framework to a few-body description, that is,  $ru_{\nu Il}^{J^{\pi}}(r)$ now describes the relative motion wave function for a two-body system. It is determined through a many-body overlap that contains all the information about the nucleon-nucleon interaction and the dynamics of the $A$-particle system.
The procedure to calculate this overlap within a symmetry-adapted framework is discussed next.

\subsection{Spectroscopic amplitudes with a symmetry-adapted (SA) basis}

We use a composite $A$-body state, calculated in the symmetry-adapted (SA) basis based on the \SU{3} symmetry \cite{Elliott_ProcRSoc245_1958,Elliott_ProcRSoc272_1962} and \SpR{3} symmetry \cite{Rosensteel_PRL38_1977,Rowe_RPP48_1985} 
 (we leave out specific details to Ref. \cite{Launey_PPNP89_2016} and references therein; see also \cite{PhysRevLett.84.1866,PhysRevC.65.054309}). The advantage is that using symmetries simplifies the calculation of the overlap without any loss of generality. In addition, we show that, if each of the clusters is described by a single equilibrium shape with suppressed vibrations, the calculations greatly simplify to a simple recursive procedure.

%To calculate the overlap in Eq. (\ref{eq:SAwithOverlap}), 
We can further refine the channels, by specifying the additional quantum numbers for the clusters as $\mathfrak{a}'={\alpha}'\omega'\kappa'(L'S')$ and $\mathfrak{a}''={\alpha}''\omega''\kappa''(L''S'')$, where $\omega' = N_{\omega}'(\lambda_{\omega}'\mu_{\omega}')$ are the \Un{3} quantum numbers with total HO excitations $N_{\omega}'$ and  \SU{3} deformation-related $(\lambda_{\omega}'\mu_{\omega}')$ (for notations, see  \cite{Launey_PPNP89_2016}). The quantity $\kappa'$ is a multiplicity in angular momentum $L'$ for a given $(\lambda'\,\mu')$, $S'$ denotes intrinsic spin, and $\alpha'$ is the set of the remaining additional quantum numbers (likewise for $\mathfrak{a}''$). 
For a single channel, the $A$-body state is projected onto two localized clusters given by a single \SU{3} configuration  (equilibrium shape) and a relative motion that is allowed any excitations; for systems that require  clusters with mixed shapes, multiple channels need to be considered. For completeness, we note that the relative motion wave function respects the \Un{3} symmetry and is described by a single \Un{3} irrep $\eta(\eta\,0)$, that is, $\chi_{\eta l} \equiv \chi_{\eta(\eta\,0) l}$.

The composite $A$-particle wave function in Eq. (\ref{eq:SAwithOverlap}) is expanded in the symplectic basis.  For example, for a single symplectic  irrep, with $\mathfrak{a}={\alpha}\sigma (LS)$ and coefficients $c$ calculated in the no-core shell model:
\begin{align}
\label{sp}
    |(A)\alpha \sigma (LS) J^{\pi}M\rangle
    &
    =
    \sum_{{n}\rho \omega \kappa }
    c_{{n} \rho \omega \kappa L }
     |{\alpha}\sigma {n} \rho \omega \kappa (L S) J^{\pi}M\rangle,
\end{align}
where $\sigma =N_{\sigma }(\lambda_{\sigma }\mu_{\sigma })$ denotes the bandhead (equilibrium shape) for this irrep; $\omega =N_{\omega }(\lambda_{\omega }\mu_{\omega })$ is given by the coupling of the bandhead $\sigma $ to a number of symplectic raising operators $\mathcal{A}^{(20)}$, symmetrically coupled to $n =N_{n}(\lambda_{n}\,\mu_{n})$. The $\rho$ and $\kappa$ are multiplicity labels [we have dropped the label ``(A)" on the right-hand side for brevity of notations]. This is generalizable to a number of symplectic irreps, however, typically a single symplectic irrep accounts for a significant portion of the wave function, often as much as $70-80\%$ \cite{DytrychLDRWRBB20,Dytrych_PRL98_2007}. The wave function for the $A$-particle system can be computed using any many-body formalism, as long as it is ultimately expanded in symplectic basis states $|{\alpha}\sigma{n} \rho \omega \kappa (LS) J^\pi M \rangle$.

Using Eq. (\ref{sp}) and through coupling of the clusters to good quantum numbers $\omega_{\rm c}\kappa_{\rm c} (L_{\rm c}S_{\rm c})I $ (with $\alpha_{\rm c}\equiv {\alpha}'\omega'S'{\alpha}''\omega''S''\rho_{\rm c}$), followed by coupling to the relative motion quantum numbers, the expression for the spectroscopic amplitude in Eq. (\ref{eq:SAwithOverlap}) can be rewritten:
\begin{widetext}
    \begin{align}
        \label{eq:specamp}
        u_{\nu Il}^{J^{\pi}}(r)
        & =
         \delta_{SS_{\rm c}}
         \sum_{\eta}
        R_{\eta l}(r)
        \sum_{\rho_{\rm c} \omega_{\rm c}\kappa_{\rm c} L_{\rm c}}
        \Pi_{L_{\rm c}S_{\rm c}I'I''I}
        \begin{Bmatrix} 
            L'  & S'  & I'  \\ 
            L'' & S'' & I'' \\ 
            L_{\rm c}   & S_{\rm c}   & I 
        \end{Bmatrix}
        \langle \omega'\kappa'L'; \omega''\kappa''L'' \| \omega_{\rm c}\kappa_{\rm c} L_{\rm c} \rangle_{\rho_{\rm c}}
        \sum_{\rho\omega\kappa L }
        \Pi_{L }
        \begin{Bmatrix} 
            S_{\rm c}  & L_{\rm c} & I   \\ 
            l  & J & L  
        \end{Bmatrix}
        \nonumber
        \\
        &\times
        (-)^{l+I+L +S_{\rm c}}
        \langle \omega_{\rm c}\kappa_{\rm c} L_{\rm c}; (\eta  0)l \| \omega \kappa L  \rangle_{\rho }
        \sum_{n=N_n(\lambda_n\,\mu_n)}
        c_{{n} \rho \omega \kappa L } \,
        \langle
        {\alpha}\sigma {n} \rho \omega \kappa LM_L 
        |
        (\alpha_{\rm c}\omega_{\rm c}; \eta(\eta\,0))\rho \omega \kappa LM_L
        \rangle,
    \end{align}
\end{widetext}
where the channel $\nu$, for a single symplectic irrep is now identified by
\begin{equation}
    \nu=\{{\alpha}\sigma S;{\alpha}'\omega'\kappa'L'S'I'^{\pi'};{\alpha}''\omega''\kappa''L''S''I''^{\pi''}\},
\end{equation}
$\Pi_{X}=\sqrt{2X+1}$, and the double-bar coefficients are \SU{3} reduced Clebsch-Gordan coefficients \cite{Draayer_JMP14_1973}. Note that the overlap 
$\langle {\alpha}\sigma {n} \rho \omega \kappa LM_L | (\alpha_{\rm c}\omega_{\rm c}; \eta(\eta\,0))\rho \omega \kappa LM_L \rangle$ 
does not depend on $M_L$
and that both cluster and symplectic states are normalized (normalization coefficients are discussed below). Calculating this overlap within a microscopic framework is non-trivial. In this study, we use
a recursive procedure, as discussed next.

We assume that the \SU{3} configuration of each of the two clusters is the bandhead  of a symplectic irrep, that is, the $\mathcal{B}^{(0\,2)}$ symplectic lowering operator annihilates it,
$\mathcal{B}^{(0\,2)} |{\alpha}'\omega'\kappa'(L'S')I'^{\pi'}M_{I'} \rangle=0$ (and similarly for the second cluster). While this condition can be relaxed if necessary (Appendix \ref{appA}), it implies that $2\ho$ excitations within the $\sigma$ symplectic irrep ($N_n \rightarrow N_n +2 $) are equivalent to  exciting the cluster relative motion by $2\ho$ ($\eta \rightarrow \eta +2$), without having any effect on the clusters. Hence, if the overlap in Eq. (\ref{eq:specamp}) is known, the overlap for  the  $\eta +2$ relative  motion can then be calculated, 
as prescribed in Ref.  \cite{Suzuki_NPA448_1986},
 through the relation:
\begin{widetext}
    \begin{align}
    \label{eq:recursion}
    \sum_{ n_{0}}
    &
    (-)^{ {\omega} -\omega_{0} }
    U[\sigma n_{0}  {\omega} (0\,2);\omega_{0} \rho_{0} 1; {n} 1 {\rho} ]
    ( {n}  \| \mathcal{B}^{(0\,2)} \| n_{0} )
    [\Delta\Omega_{K}(n_{0} ,\omega_{0} ; {n} , {\omega} )]^{1/2}\,\,
    \langle{\alpha}\sigma n_{0} \rho_{0} \omega_{0} \Lambda_{0}  | (\alpha_{\rm c} \omega_{\rm c};\eta_0(\eta_0\,0))\rho_{0} \omega_{0} \Lambda_{0} \rangle
    \nonumber
    \\
    &
    =
    \sqrt{\mathrm{dim}(\eta_0\,\,0)}
    ~
    U[\omega_{\rm c}(\eta_0\,\,0) {\omega} (0\,2);\omega_0 \rho_0 1;(\eta\,0)1 {\rho} ] \,\,
    \langle{\alpha}\sigma  {n}  {\rho}  {\omega}  {\Lambda}  | (\alpha_{\rm c}\omega_{\rm c};\eta(\eta \,0)) {\rho}  {\omega}  {\Lambda}\rangle,
    \\
    &({\rm with}\, n_{0}=N_{n_0}(\lambda_{n_0}\,\mu_{n_0}),\,\, N_{n_0}=N_n +2,\,{\rm and}\,\eta_0=\eta+2) \nonumber
    \end{align}
\end{widetext}
for each $\rho_0\omega_0\Lambda_0$ 
(using $\Lambda \equiv \kappa LM_{L}$).
Note that all $\rho_0\omega_0\Lambda_0$ configurations are known for a given $N_{n_0}$ within the $\sigma$ symplectic irrep. This recursion is defined for  a base case $n=0(0\,0)$ (implying $N_n=0$, $\rho=1$, and $\omega= \sigma$)
\begin{equation}
\label{rec_bc}
\langle{\alpha}\sigma  {n=0(0\,0)}  {\rho}  \omega  {\Lambda}  | (\alpha_{\rm c}\omega_{\rm c};\eta_{\mathrm{min}}(\eta_{\mathrm{min}} \,0)) {\rho}  {\omega}  {\Lambda}\rangle=1,
\end{equation}
with $\eta_{\mathrm{min}}=N_{\sigma}-(N_{\omega}'+N_{\omega}''-3/2)$, where $N_{\omega}'$ and $N_{\omega}''$ are the total HO quanta for the clusters, and $N_{\sigma}$ is defined as the total HO quanta minus 3/2 to remove the spurious CM motion in the symplectic state \cite{Tobin_PRC89_2014}.
In Eq. (\ref{eq:recursion}), the $U[\dots]$ symbol is an \SU{3} Racah coefficient \cite{Hecht_NP62_1965}, analogous to the $6$-$j$ symbol for \SU{2}. The dimension of $(\lambda\,\mu)$ is denoted with $\mathrm{dim}(\lambda\,\mu)=\half(\lambda+1)(\mu+1)(\lambda+\mu+2)$. Although we retain $\Lambda_{0}$ and $ {\Lambda}$ in the overlaps, it should be noted that the overlaps are independent of these labels. 

Note that for given $N_{n_0}\rho_0\omega_0\Lambda_0$, there could be several configurations that differ by $(\lambda_{n_0}\,\mu_{n_0})$, that is,  $|\sigma  N_{n_0} (\lambda_{n_0}^{(1)}\,\mu_{n_0}^{(1)})  {\rho_0}  {\omega_0}  {\Lambda_0} \rangle$, $|\sigma  N_{n_0} (\lambda_{n_0}^{(2)}\,\mu_{n_0}^{(2)})  {\rho_0}  {\omega_0}  {\Lambda_0} \rangle$
, $\dots$, while in some cases there is a single $(\lambda_{n_0}\,\mu_{n_0})$ and Eq. (\ref{eq:recursion}) becomes a simple recursive formula. These cases are of a special interest, and will be the ones considered in this study.

Both symplectic and cluster states in Eq. (\ref{eq:recursion}) are normalized. The symplectic states are normalized through the use of the $\mathcal{K}$-matrix \cite{Rowe_RPP48_1985, Rowe_JPALettEd17_1984, Rowe_JMP25_1984}. Although the $\mathcal{K}$-matrix is not diagonal in general, in the limit of large $\sigma$, it reduces to diagonal \cite{Rowe_JPALettEd17_1984,Hecht_JPA18_1985}. For the diagonal matrix element, we adopt the notation
\begin{align}
    \label{kmat1}
    \Delta\Omega_{K}(n_{0},\omega_{0};n,\omega)
    =\Omega_{K}(n_{0},\omega_{0}) - \Omega_{K}(n,\omega),
\end{align}
with $\mathcal{K}$-matrix coefficients given by \cite{Rowe_RPP48_1985}
\begin{align}
    \label{kmat2}
    \Omega_{K}&(n,\omega)
    \\
    &
    =
    \frac{1}{4}\sum_{j=1}^{3}\left[ 2\omega_{j}^{2} -n_{j}^{2}+8(w_{j}-n_{j})-2j(2w_{j}-n_{j})\right],     \nonumber
\end{align}
where $n_{1}=\frac{N_{n}+2\lambda_{n}+\mu_{n}}{3}$, $n_{2}=\frac{N_{n}-\lambda_{n}+\mu_{n}}{3}$, and $\frac{N_{n}-\lambda_{n}-2\mu_{n}}{3}$ (the $\omega_{j}$ are similarly defined for $N_{\omega},\lambda_{\omega},\mu_{\omega}$). The normalization for the cluster states has been previously derived. For cluster systems comprised of an $\alpha$ particle and a heavy fragment, with total particle number $12\leq A \leq24$, the normalization is available in Ref. \cite{Hecht_NPA313_1979}. 
Other selected cases are available in \cite{Hecht_NPA356_1981}. 

The antisymmetrization of the cluster wave function can be straightforwardly calculated in the overlap in Eq. (\ref{eq:SAwithOverlap}) since the $A$-particle wave function is already antisymmetrized, yielding a factor that depends only on $A-a$ and $a$. Since this factor remains the same for each $N_{n_0}$, it propagates through the recursive procedure down to the base case and is absorbed in the overlap (\ref{rec_bc}).

%%%%%%%%%%%%%%%%%%%%
%%%%% Results %%%%%%
%%%%%%%%%%%%%%%%%%%%
\section{Results}
The formalism is demonstrated for  \nuc{20}{Ne}, which is a well-known cluster system. In particular, the excited $1^{-}$ state, 5.79 MeV above the ground state, is understood to be close to a pure $\alpha$-cluster state. This level is 1.06 MeV above the $\alpha+\nuc{16}{O}$ threshold, which corresponds to the energy in the CM frame of the cluster system ($E$), or 1.33 MeV laboratory kinetic energy of $\alpha$,  given by 
 $E_{\mathrm{lab}}=(\frac{m_{^{16} \rm O}+m_{\alpha}}{m_{^{16} \rm O}})E$.

In this study, the $A$-particle wave function is calculated using two many-body frameworks: (1) the microscopic NCSpM \cite{Dreyfuss_PLB727_2013}, and (2) the $\abinitio$ SA-NCSM \cite{Launey_PPNP89_2016}. Both models have yielded energy spectra and observables (radii, quadrupole moments, and $E2$ transitions) in close agreement with experiment \cite{Dreyfuss_PRC95_2017, Tobin_PRC89_2014, Launey_PPNP89_2016, DytrychLDRWRBB20}.
  
The NCSpM uses the symplectic \SpR{3} basis and a microscopic many-nucleon Hamiltonian that includes the nucleon-nucleon (NN) long-range part, expressed in terms of the $Q\cdot Q$ interaction, and a spin-orbit interaction (for details, see \cite{Dreyfuss_PLB727_2013, Tobin_PRC89_2014}). Wave functions are calculated in the basis of a single symplectic irrep, including all excitations above the lowest-energy configuration up to the truncation parameter $\Nmax$, whereas the mixing between several symplectic irreps is included by solving the SA-NCSM in small model spaces \cite{Tobin_PRC89_2014,Dreyfuss_PLB727_2013,Dreyfuss_PRC95_2017}. The $\abinitio$ SA-NCSM is a no-core shell model using realistic NN interactions within either the \SU{3} symmetry-adapted basis \cite{Dytrych_PRL111_2013} or the \SpR{3} symmetry-adapted basis \cite{Launey_PPNP89_2016,DytrychLDRWRBB20}. In this study, we use the  NNLO$_{\rm opt}$ chiral potential \cite{Ekstrom_PRL110_2013}. To distinguish between the two bases, the results presented here will be denoted as  ``SA-NCSM/\SU{3}" and ``SA-NCSM/\SpR{3}", respectively. For the \SU{3} basis, we make the approximation that a given \SU{3} basis state belongs to only a single symplectic irrep, ignoring a small mixing of additional symplectic irreps, the validity of which is discussed below.
\begin{figure}[t]
    \centering
    \includegraphics[width=0.47\textwidth]{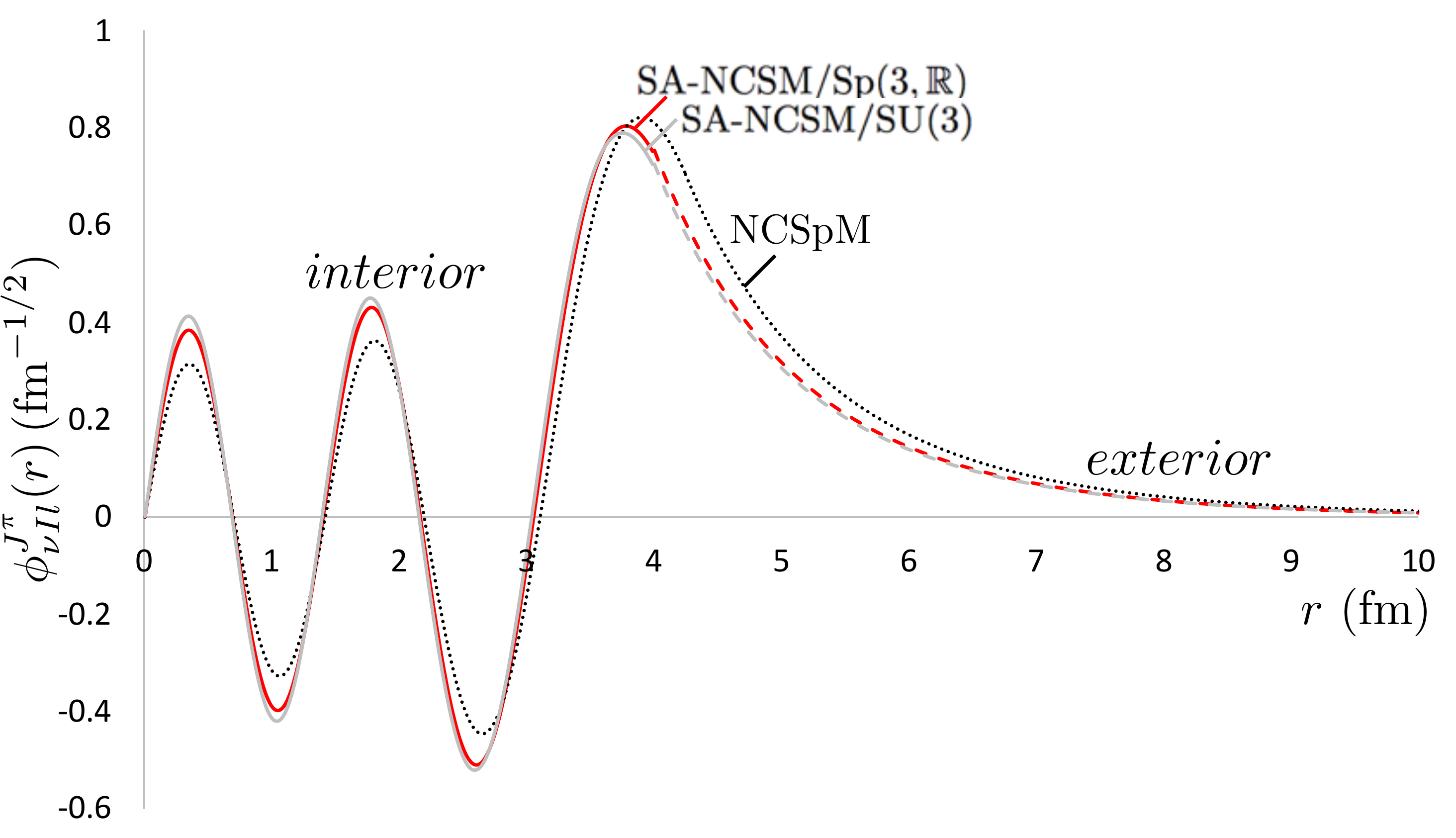}
    \caption
    {
        \label{fig:sympvsSU3_0}
Wave function of the $l=0$ $\alpha+\nuc{16}{O}$ for the \nuc{20}{Ne} \gs~($\hbar\Omega=15$ MeV), calculated in the $\abinitio$ $\Nmax=8$ SA-NCSM using an \SpR{3} basis (red, labeled as ``SA-NCSM/\SpR{3}") and an \SU{3} basis (grey, labeled as ``SA-NCSM/\SU{3}"),  and in the $\Nmax=22$ NCSpM (dotted black).
Results are reported for a single symplectic  irrep, $\sigma=48.5(8\,0)$  (see text for details).
The interior and exterior contributions to the wave function are indicated with solid and dashed lines, respectively.
    }
\end{figure}
\begin{figure*}[th]
    \subfloat[\label{fig:sympvsSU3_1}]{%
      \includegraphics[width=.49\linewidth]{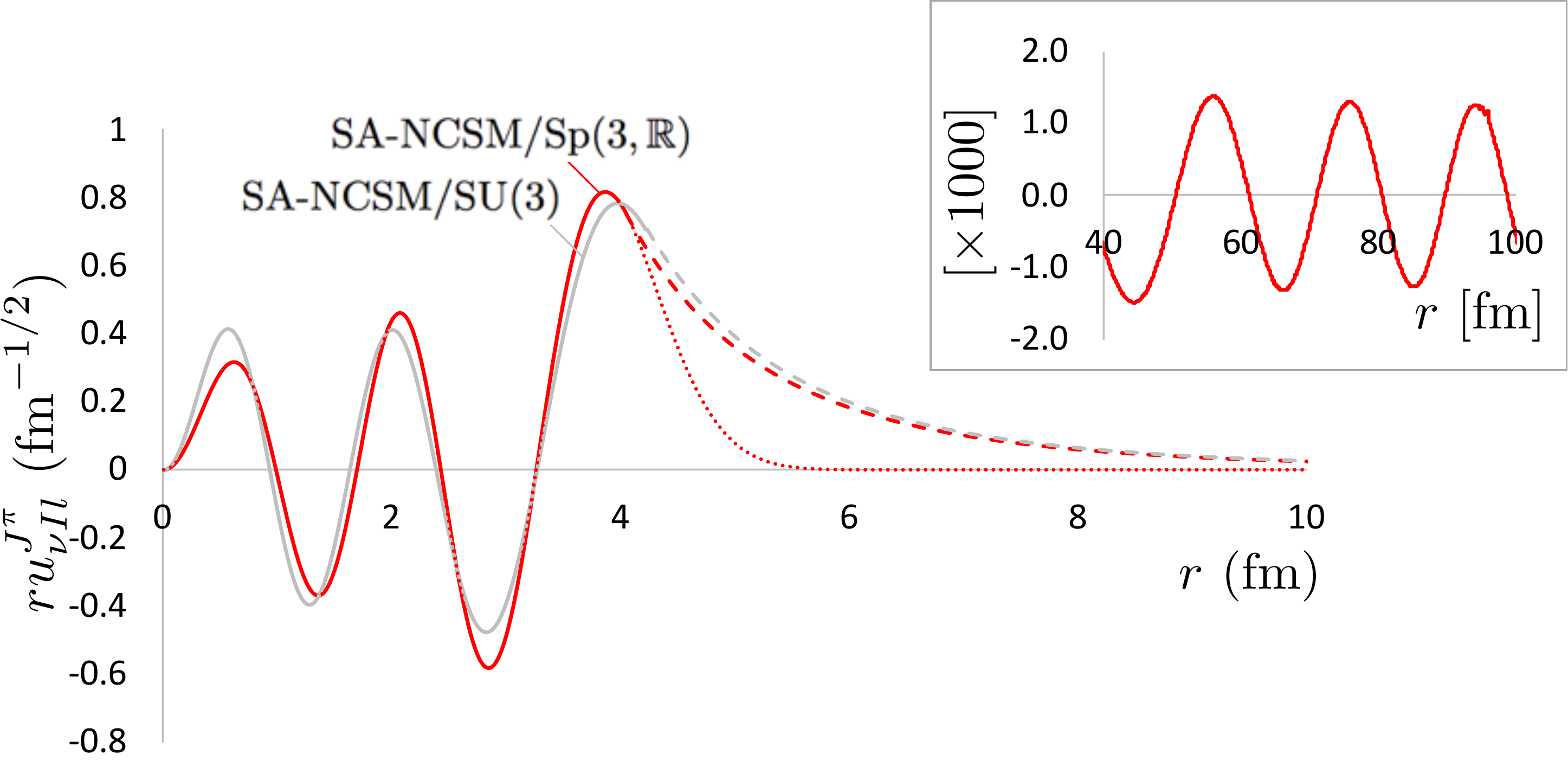}%
    }\hfill
    \subfloat[\label{fig:SA_MSparams}]{%
      \includegraphics[width=.49\linewidth]{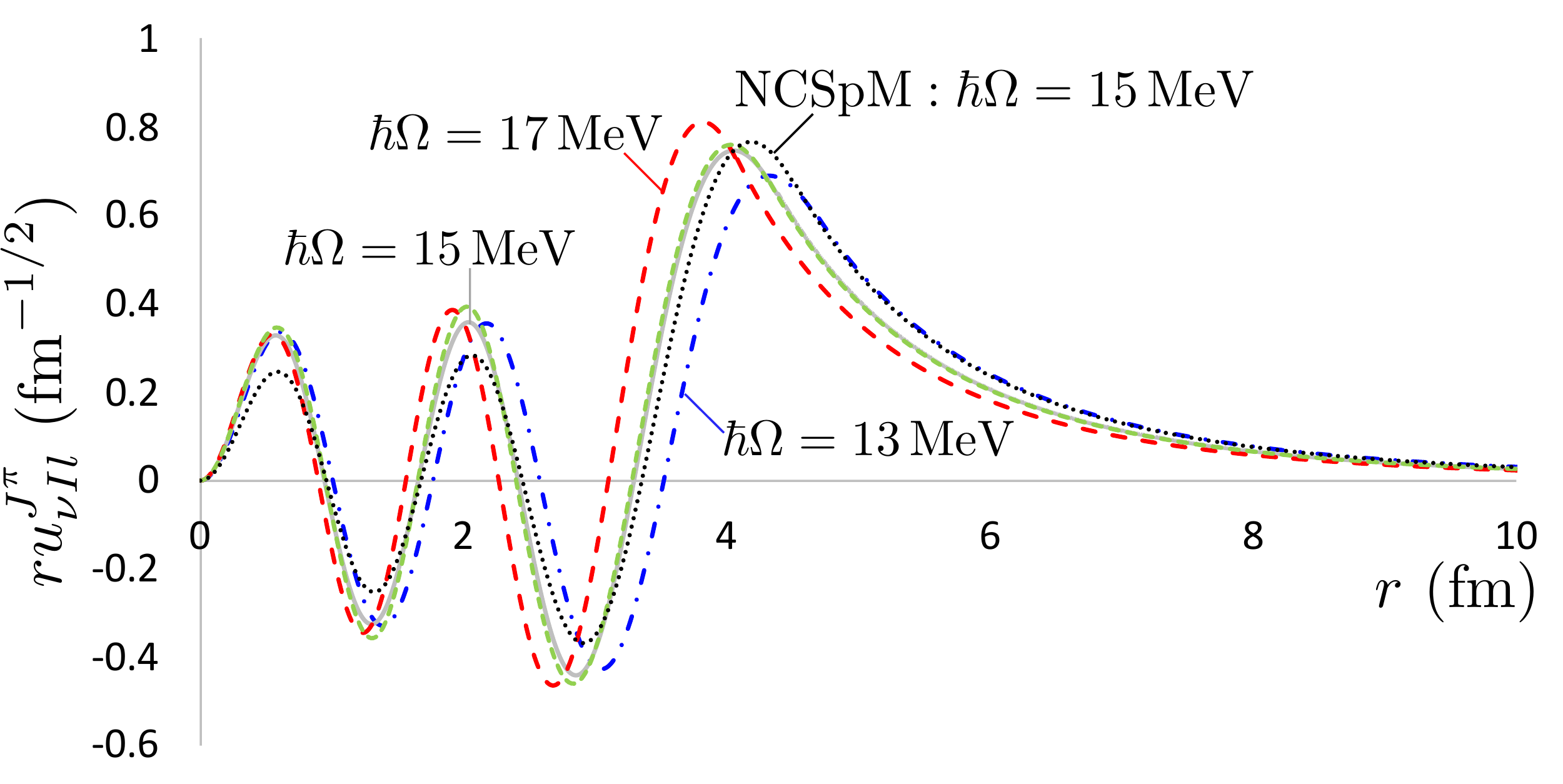}%
    }
    \caption{Spectroscopic amplitudes for the $l=1$ $\alpha+\nuc{16}{O}$ at energy $E=1.06$ MeV, calculated from the lowest $1^{-}$ state in $\nuc{20}{Ne}$. Results are reported for a single symplectic  irrep, $\sigma=49.5(9\,0)$  (see text for details). (a) The \nuc{20}{Ne} wave functions are calculated using the $\abinitio$ SA-NCSM [with \SU{3} basis (grey) and \SpR{3} basis (red)] with $\Nmax=5$ and $\hbar\Omega=15$ MeV. Interior and exterior (real part) components are shown with solid and dashed lines, respectively. For the \SpR{3} case, we show the spectroscopic amplitude without matching to the Coulomb-Hankel functions (dotted red), and the asymptotic oscillatory behavior of the real part of the wave functions (inset). (b) The composite \nuc{20}{Ne} wave functions are calculated using the SA-NCSM/\SU{3} with $\Nmax=9$ [$\hbar\Omega=13$ MeV (dash-dotted blue) and $15$ MeV (solid grey)], $\Nmax=7$ [$\hbar\Omega=15$ MeV (small-dash green) and $17$ MeV (large-dash red)], and NCSpM with $\Nmax=23,\,\hbar\Omega=15$ MeV (dotted black).
    }
    \label{fig:specampsall}
\end{figure*}

We have shown that for the \nuc{20}{Ne} \gs,  calculated in the $\abinitio$ SA-NCSM with $\Nmax=8$ and $\hbar\Omega=15$ MeV, $~70\%$ of the wave function is described with just the leading $\sigma=N_{\sigma}(\lambda_{\sigma}\mu_{\sigma})=48.5(8\,0)$ symplectic irrep \cite{DytrychLDRWRBB20}. In addition, the wave functions show that there are a few states within the irrep that typically contribute the most (see also \cite{Tobin_PRC89_2014}), so that we can consider only the stretched-coupled states, which are the most deformed states 
and the next most deformed states in the irrep. Namely, for a given $N_{n}$, the most deformed states are
$(\lambda_{n},\,\mu_{n})=(N_{n},\,0)$ and $(\lambda_{\omega},\,\mu_{\omega})=(\lambda_{\sigma}+N_{n},\,\mu_{\sigma})$, and the next most deformed states are 
$(\lambda_{n},\, \mu_{n})=(N_{n}-4,\,2)$ and $(\lambda_{\omega},\, \mu_{\omega})=(\lambda_{\sigma}+N_{n}-4,\,\mu_{\sigma}+2)$.
By keeping only the most deformed and second most-deformed contributions to the wave function, we retain $99\%$ of that irrep (the wave functions can be easily renormalized to ensure a norm of 1). In the present study, this set of contributions is used for all \nuc{20}{Ne} wave functions, properly normalized to 1, which, in turn, simplifies the overlap calculations  (\ref{eq:recursion}) to a recursive procedure.

From these many-body calculations, we determine the coefficients $c_{n\rho\omega\kappa L}$ in Eq. (\ref{eq:specamp}), while the overlap is calculated using Eq. (\ref{eq:recursion}) -- this yields the spectroscopic amplitudes $u_{\nu Il}^{J^{\pi}}(r)$, which determine ANCs and alpha widths.  The channels $\nu$ are defined for  $\omega'=36(0\,0)$ and  $I'^{\pi'}=0^+$ for $^{16}$O, and $\omega''=6(0\,0)$ and  $I''^{\pi''}=0^+$  for $^4$He,  implying that $\omega_c=40.5(0\,0)$ and $\eta_{\min}=8$ for the $0^+_{\rm gs}$, $2^+_1$ and $4^+_1$ states in $^{20}$Ne (or $\eta_{\min}=9$ for the $1^-_1$ state). For this case, the norm of the cluster sate is taken from Ref. \cite{Hecht_NPA313_1979}.

Although alpha partial widths for states close to the threshold are not generally directly measurable in experiment because of very low cross sections at astrophysically relevant energies, the natural width of the \nuc{20}{Ne} $1^{-}$ state is known to be $28(3)$ eV \cite{PhysRevC.22.356,Constantini_PRC82_2010}. There is a nearby $3^{-}$ state at 5.62 MeV above the ground state (0.89 MeV above the $\alpha+\nuc{16}{O}$ threshold) that should typically be treated in a coupled-channels framework with the $1^{-}$ state. However, both states are very narrow, especially the $3^{-}$ state, and, as their coupling is negligible, they can be treated in separate single-channel calculations. Calculations for the $3^{-}$ state are left for another study.

In all calculations shown here, we take the experimental energy for the threshold,
resonance  and 
  the bound states. A self-consistent calculation of the energy requires a number of large-scale calculations of all three systems (the composite system, the $A-a$ cluster, and the $a$ cluster) to determine converged thresholds and energy differences for each of the two clusters, which is outside the scope of this work.

 \subsection{Bound state wave functions and resonance spectroscopic amplitudes}
\label{subsec:SA_BS}
 
We study the cluster structure of the \nuc{20}{Ne} ground state, using the $R$-matrix theory, that is, Eqs. (\ref{eq:ext})-(\ref{eq:int}) and matching near the effective range of the interaction (Fig. \ref{fig:sympvsSU3_0}). For these figures, $r_{c}$ is chosen so that the calculated ANC is maximized, which  coincides with convergent results for the SA-NCSM, as discussed in Section \ref{subsec:Apartwide}. Note that the ANCs (\ref{eq:ANC}) require that the internal wave function is normalized to 1 in $[0,r_c)$, which we impose on the $u_{\nu Il}^{J^{\pi}}(r)$ calculated in Eq. (\ref{eq:specamp}). The results obtained using the \SU{3} basis to compute the \nuc{20}{Ne} \gs~are nearly indistinguishable from the results using the \SpR{3} basis with a single symplectic irrep for the same $\Nmax=8$ and $\hbar\Omega=15$ MeV. This implies that the predominant contribution to each \SU{3} state is indeed coming from the leading symplectic irrep, hence, this assumption used in the \SU{3}-based calculations is reasonable. To guide the eye, we compare to the NCSpM wave function in larger model spaces $\Nmax=22$ and with $\hbar\Omega=15$ MeV (Fig. \ref{fig:sympvsSU3_0}). The result is very similar to the SA-NCSM calculations, suggesting that the bound-state physics is reasonably described already within $\Nmax=8$ model spaces.

\begin{figure}[th]
    \subfloat[$\hbar\Omega=13$ MeV\label{fig:hw13width}]{%
      \includegraphics[width=.45\textwidth]{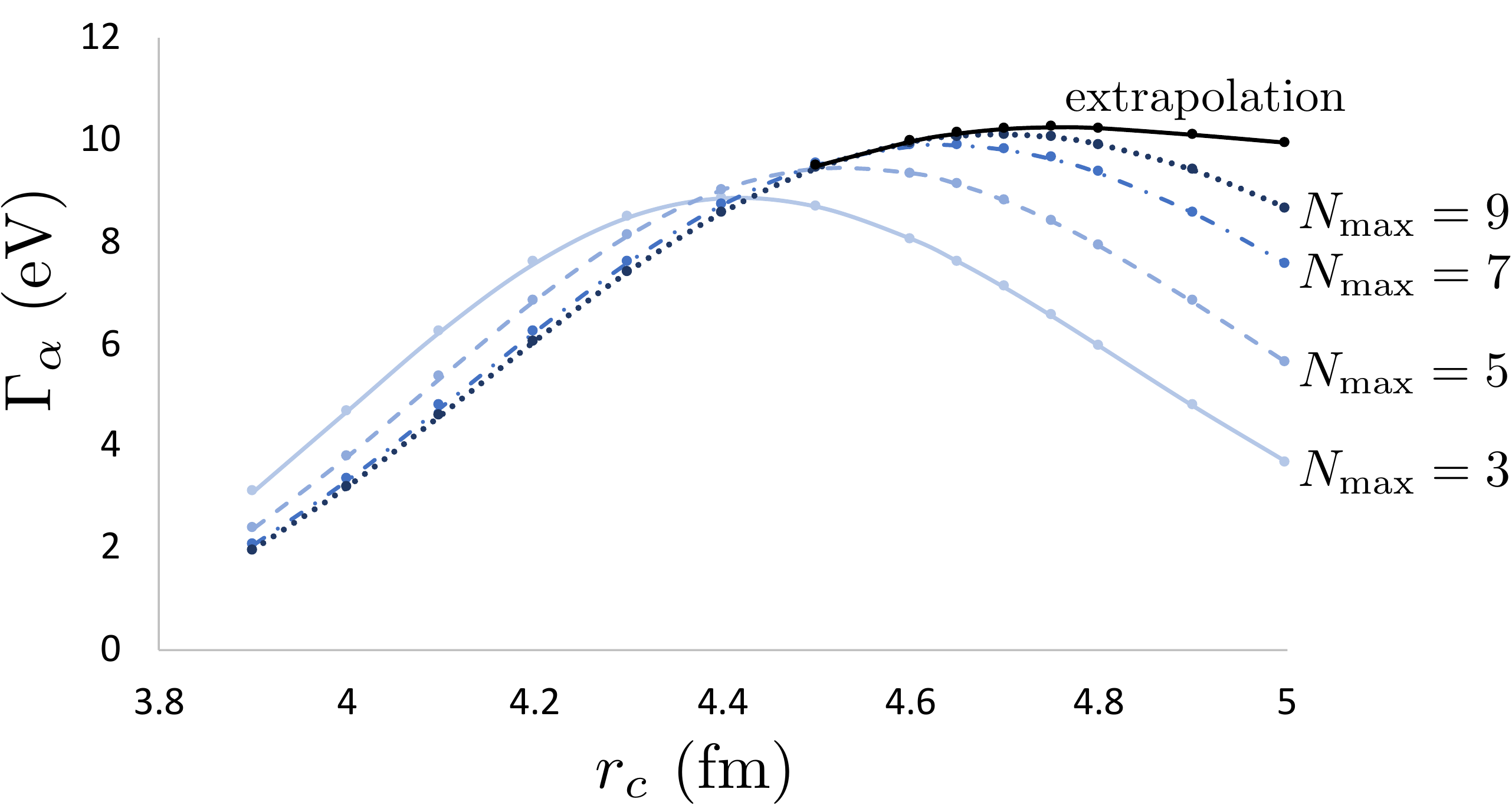}%
    }\\
    \subfloat[$\hbar\Omega=15$ MeV\label{fig:hw15width}]{%
      \includegraphics[width=.45\textwidth]{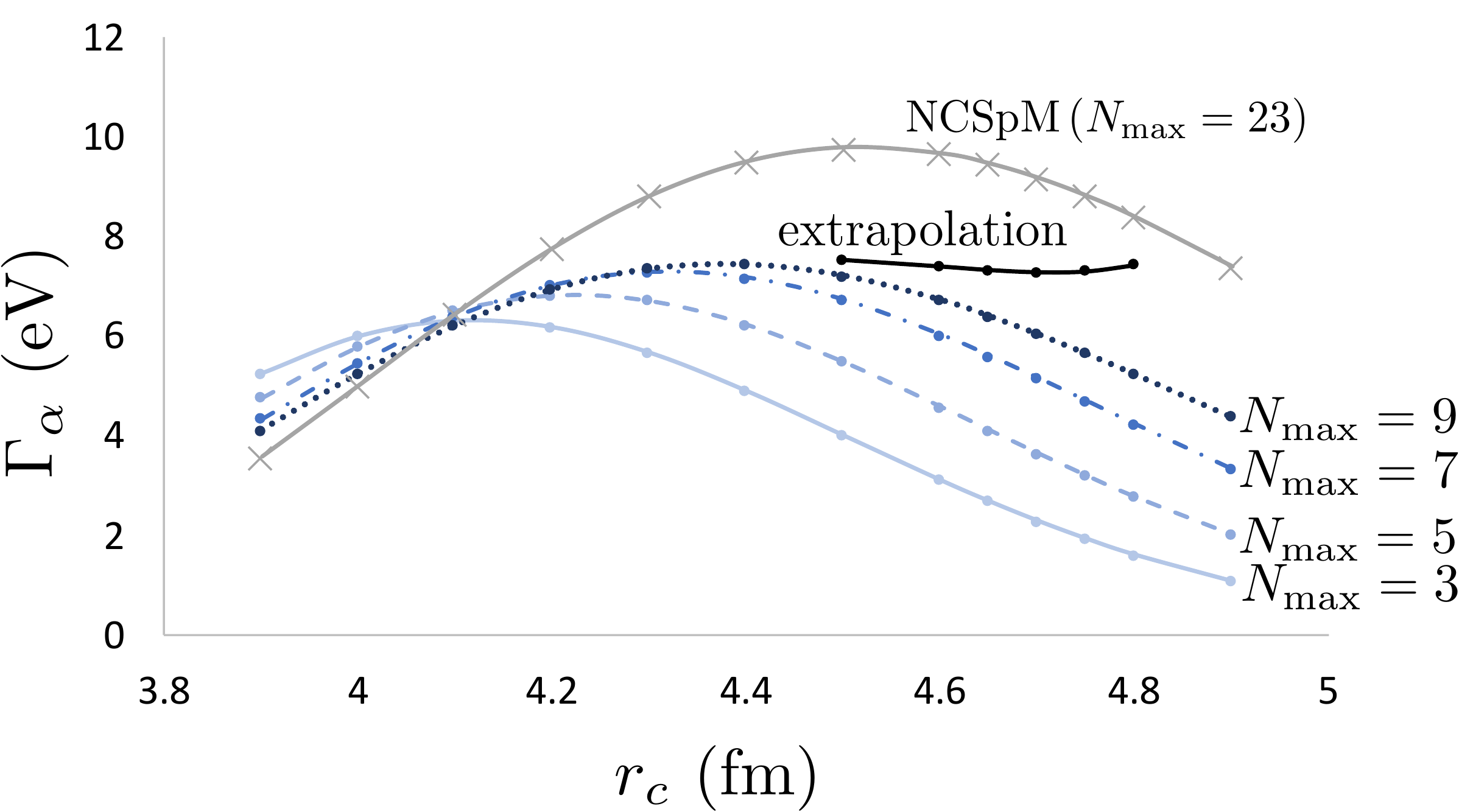}%
    }\\
    \subfloat[$\hbar\Omega=17$ MeV\label{fig:hw17width}]{%
    \includegraphics[width=.45\textwidth]{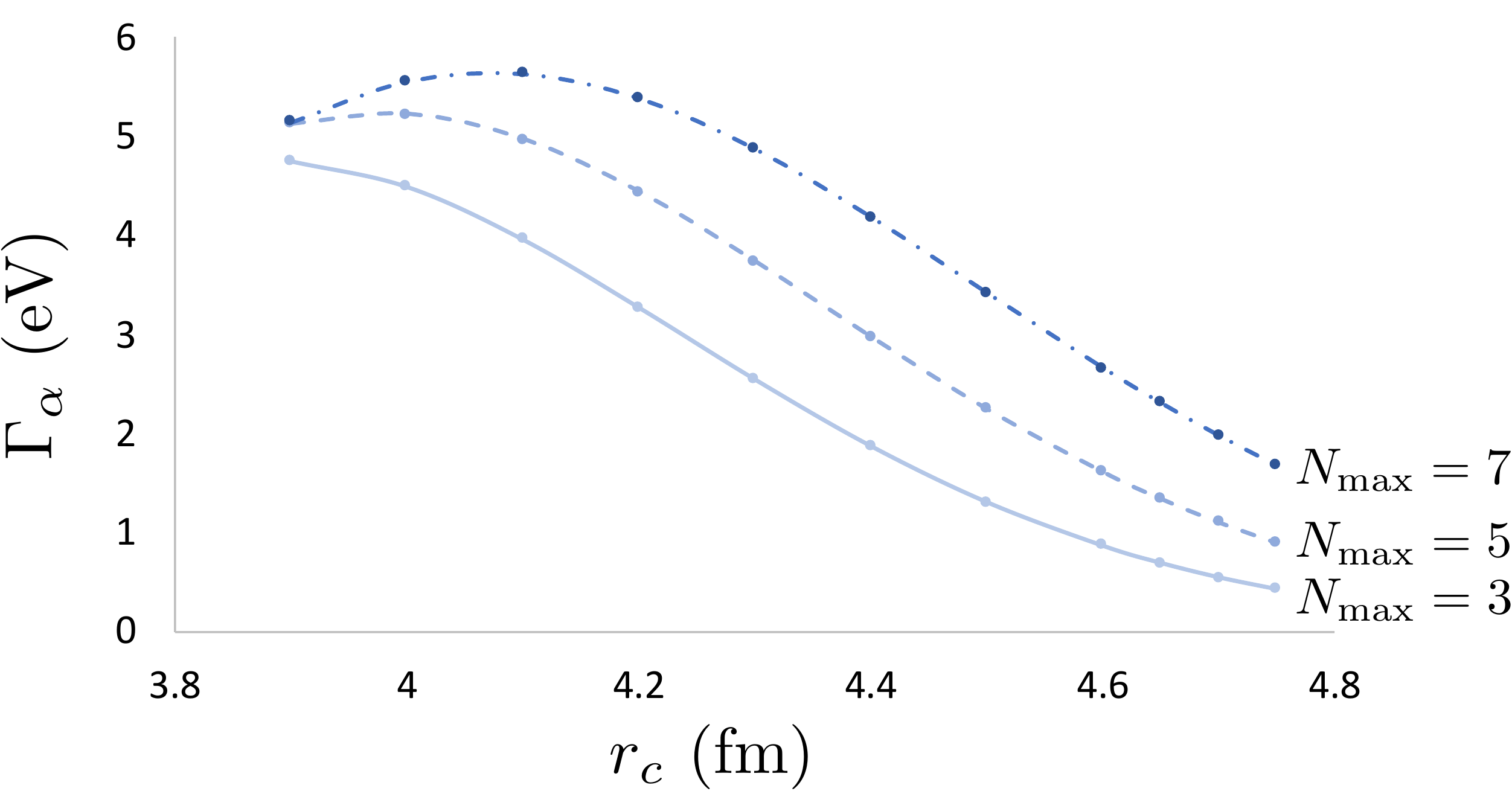}%
  }
    \caption{Alpha partial width $\Gamma_{\alpha}$ as a function of the channel radius $r_{c}$  for the $1^{-}$ resonance at 1.06 MeV (relative to the $\nuc{16}{O}+\alpha$ threshold) in \nuc{20}{Ne}. The \nuc{20}{Ne} wave functions are calculated using the $\abinitio$ SA-NCSM [with \SU{3} basis] (blue) and NCSpM (grey $\times$) for increasing $\Nmax$ model spaces and with (a) $\hbar\Omega=13$ MeV, (b) $\hbar\Omega=15$ MeV, and (c) $\hbar\Omega=17$ MeV. The extrapolation (black) is determined with Shanks transformation on the $\Nmax=5,7,$ and 9 data.
    }
    \label{fig:widthsextrap}
\end{figure}

In a similar fashion, we study the $1^{-}$ resonance in $\nuc{20}{Ne}$ near the $\alpha+\nuc{16}{O}$ threshold (Fig. \ref{fig:sympvsSU3_1}). Again, matching is done for a value of $r_{c}$ that yields a maximum partial width $\Gamma_{\alpha}$, which coincides with convergent results for the SA-NCSM, as discussed in Section \ref{subsec:Apartwide}. The \nuc{20}{Ne} $1^{-}$ wave function is calculated in the $\abinitio$ SA-NCSM with $\Nmax=5$ and $\hbar\Omega=15$ MeV, where the symplectic-based results 
%based on a symplectic basis 
use the leading $\sigma=49.5(9\,0)$ irrep. The close agreement between the ``SA-NCSM/\SU{3}" and ``SA-NCSM/\SpR{3}" spectroscopic amplitudes further confirms that the assumption used in the \SU{3}-based calculations is reasonable. 

We note that, because the $1^{-}$ state in \nuc{20}{Ne} is a resonance, the wave function is not expected to decay in the asymptotic regime as it does in the \gs~case; rather, the exterior resonance wave function has oscillatory behavior, as shown in the inset in Fig. \ref{fig:sympvsSU3_1}. In this exterior regime, the form of the spectroscopic amplitude is determined by the spherical Coulomb-Hankel functions $H_{l}^{+}(\eta,kr)$. Without matching to the Coulomb-Hankel functions, the spectroscopic amplitude is significantly changed (see Fig. \ref{fig:sympvsSU3_1}, red dotted curve). Matching, therefore, is of integral importance to obtaining the correct asymptotic behavior, whereas many-body descriptions that properly account for spatially extended configurations are key to obtaining the interior behavior, which, in turn, determines widths. 

Further, we study the dependence on the model parameters, $\Nmax$ and $\hbar\Omega$  (Fig. \ref{fig:SA_MSparams}). E.g., for $\hbar\Omega=15$ MeV, the spectroscopic amplitudes for $\Nmax=7$ and $\Nmax=9$ are nearly indistinguishable (see the solid grey and dashed green curves in Fig. \ref{fig:SA_MSparams}). In addition, while the results for the largest model space under consideration exhibit some $\hbar\Omega$ dependence, the spread is not significant.
 The main effect is that the surface peaks are slightly shifted toward  larger separation distances for smaller $\hbar\Omega$ values.
 To guide the eye, we again include the  spectroscopic amplitude computed using the \nuc{20}{Ne} NCSpM $1^{-}$ wave function in larger model spaces $\Nmax=23$ and with  $\hbar\Omega=15$ MeV (Fig. \ref{fig:SA_MSparams}, black dotted curve). The NCSpM spectroscopic amplitude is matched at a larger radius (meaning the associated $\Gamma_{\alpha}$ is maximized for a larger radius), compared to the $\abinitio$ SA-NCSM/\SU{3} $\hbar\Omega=15$ MeV spectroscopic amplitudes. Nonetheless, it is interesting that the NCSpM tail coincides with the one obtained from the $\hbar\Omega=13$ MeV SA-NCSM. All of the spectroscopic amplitudes in Fig. \ref{fig:SA_MSparams} have small inner peaks, compared to a relative large and broad surface 
peak (the largest peak on the right), and these features are slightly more exaggerated in the NCSpM spectroscopic amplitude. 
In comparison to the \nuc{20}{Ne} ground state (Fig. \ref{fig:sympvsSU3_0}) for the same $\hbar\Omega$, the $1^{-}$  resonance  exhibits slightly smaller  inner peaks and wider surface peak associated with surface clustering, in agreement with cluster model outcomes \cite{KanadaEnyo_PTEP2014_2014,Suzuki_KDLbook}. 

The shapes of the cluster wave functions for the ground state and $1^-$ state (Fig. \ref{fig:sympvsSU3_0} and Fig. \ref{fig:SA_MSparams}), including the asymptotics, agree reasonably well with GCM cluster wave functions \cite{KanadaEnyo_PTEP2014_2014}, and even models that have been specifically tailored to describe cluster states \cite{Buck_PRC11_1975}. E.g., in comparison to the results in Ref. \cite{KanadaEnyo_PTEP2014_2014}, the $\hbar\Omega=13$ MeV results exhibit only slight differences that consist of  larger inner probabilities and peaks shifted to larger separations distances, as well as a longer  tail of the $1^-$ resonance. This suggests that the shell-model wave functions are able to describe the spatially extended clusters, while providing a slightly larger spatial overlap of  the two clusters due to shell-model configuration mixing.

Although a large spectroscopic factor cannot be used as the sole indicator of clustering \cite{Suzuki_KDLbook}, for completeness we report spectroscopic factors for the $1^{-}$ resonance of \nuc{20}{Ne}, using Eq. (\ref{eq:SF}), for each of the spectroscopic amplitudes  in  the largest SA-NCSM model spaces shown in Fig. \ref{fig:SA_MSparams}. We calculate  spectroscopic factors of  $SF=0.73$, $0.76$, and $0.80$ for $\hbar\Omega=13$, $15$, and $17$, MeV respectively; the SA-NCSM estimate for $\hbar\Omega=13$ MeV also agrees with the NCSpM spectroscopic factor. These values are larger than a spectroscopic factor of $SF<0.344$, calculated using a simple cluster wave function \cite{Horiuchi_PTP49_1973}.

\begin{figure}[th]
    \subfloat[$\hbar\Omega=13$ MeV\label{fig:hw13anc}]{%
      \includegraphics[width=.45\textwidth]{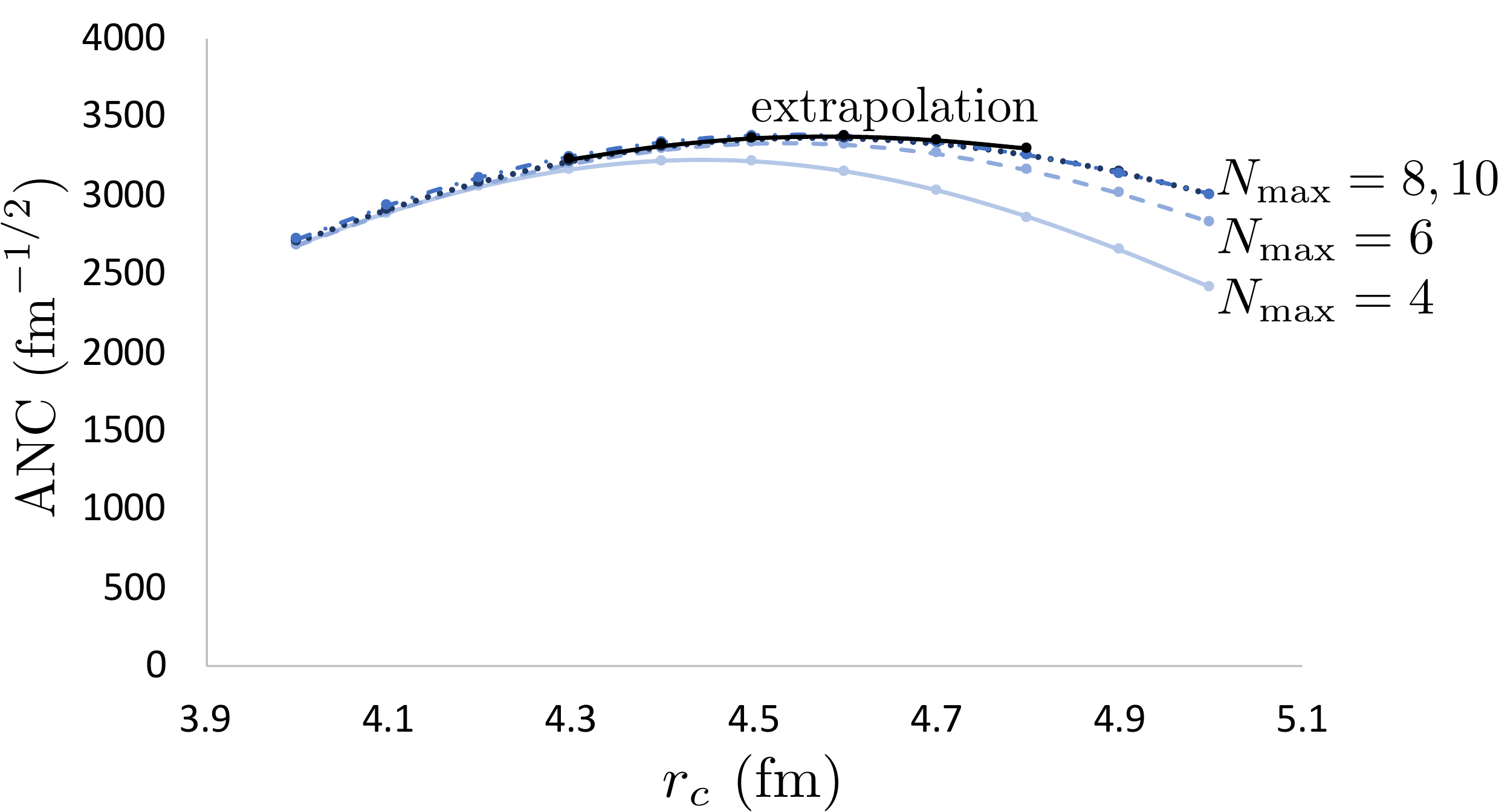}%
    }\\
    \subfloat[$\hbar\Omega=15$ MeV\label{fig:hw15anc}]{%
      \includegraphics[width=.45\textwidth]{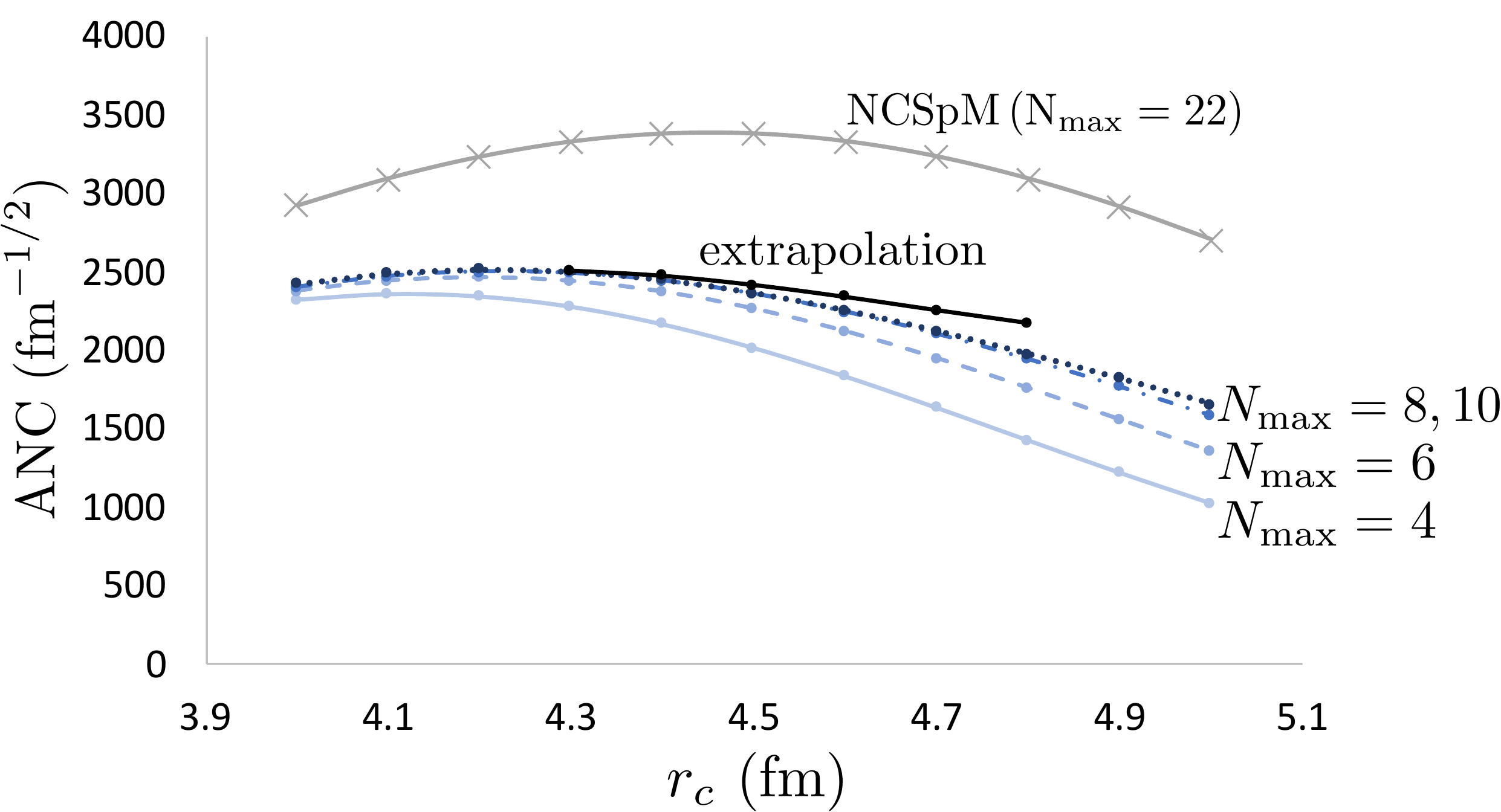}%
    }\\
    \subfloat[$\hbar\Omega=17$ MeV\label{fig:hw17anc}]{%
      \includegraphics[width=.45\textwidth]{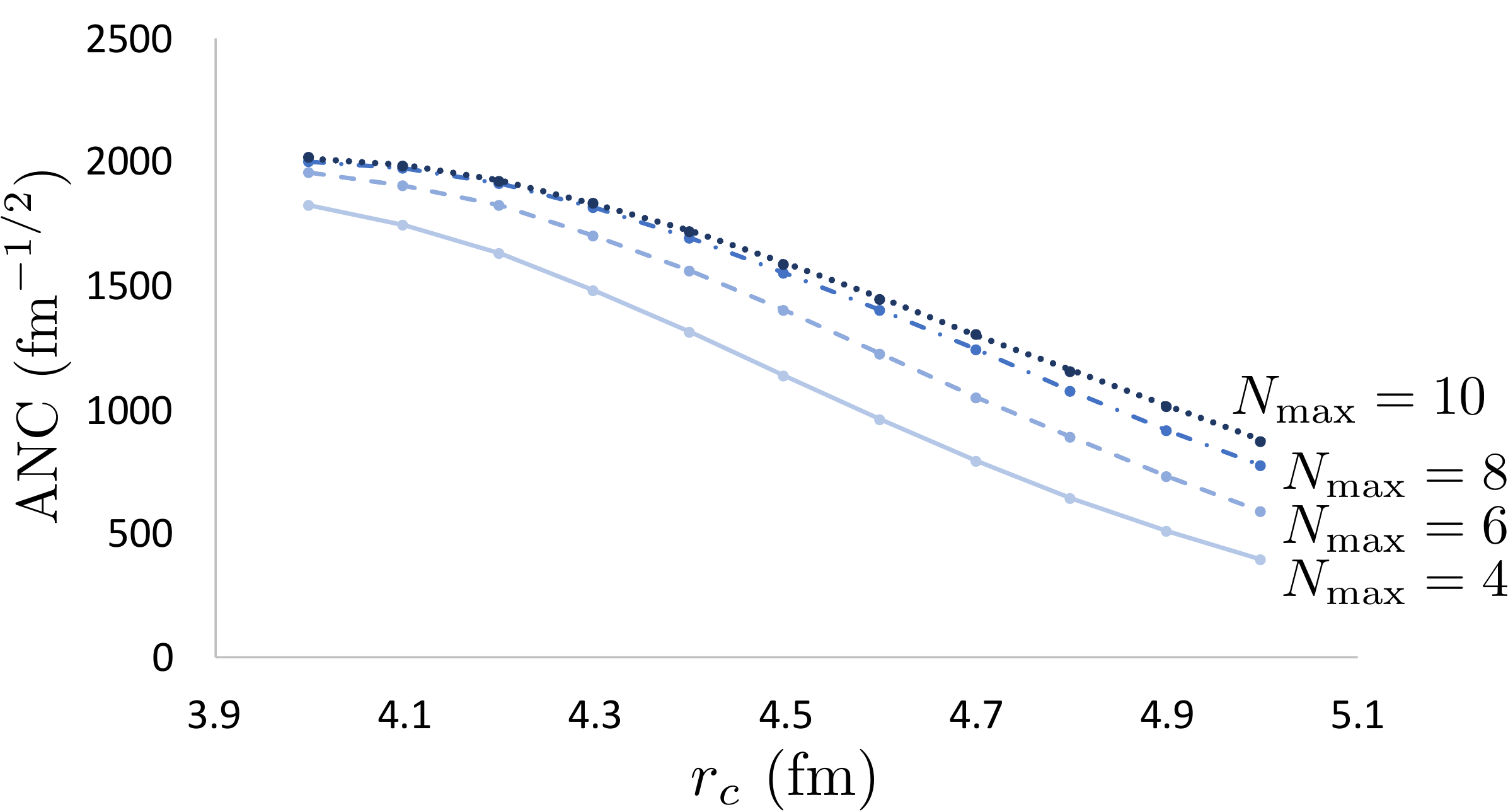}%
    }
    \caption{ANC ($C_{0}$) for the \nuc{20}{Ne} \gs~in the $l=0$ $\nuc{16}{O}+\alpha$ channel as a function of the channel radius $r_{c}$. The \nuc{20}{Ne} ground state is calculated using  the $\abinitio$ SA-NCSM [with \SU{3} basis] (blue)  and NCSpM (grey $\times$) for increasing $\Nmax$ model spaces and with (a) $\hbar\Omega=13$ MeV, (b) $\hbar\Omega=15$ MeV, and (c) $\hbar\Omega=17$ MeV. The extrapolations (black) are determined with the Shanks transformation  on the $\Nmax=6,8,$ and 10 data.
    }
    \label{fig:ANC}
\end{figure}

\subsection{$\alpha$ partial widths and ANCs}
\label{subsec:Apartwide}
The alpha partial widths $\Gamma_{\alpha}$ are determined using Eqs. (\ref{eq:width}) and (\ref{eq:specamp}). In general, the partial widths depend on the model space parameters ($\Nmax$ and \ho) through the use of the \abinitio SA-NCSM   \nuc{20}{Ne}  wave functions  in Eq. (\ref{eq:specamp}), while the matching to the exterior wave function introduces a dependence on the channel radius $r_c$. For the 1.06-MeV $1^{-}$ resonance, we find that the partial width strongly depends on  \ho~and the channel radius in small model spaces, whereas the dependence tends to decrease for increasing $\Nmax$  (Fig. \ref{fig:widthsextrap}). In particular, the $\hbar\Omega=13$ MeV results in Fig. \ref{fig:hw13width} show a clear pattern toward convergence, and for $\Nmax=7$ and $\Nmax=9$, the partial width is nearly independent of both $\Nmax$ and the channel radius for $r_{c}=4.6$-$5$ fm. Similarly, the $\hbar\Omega=15$ MeV results (Fig. \ref{fig:hw15width}) are on a convergence trend but indicate that larger model spaces are needed. In contrast, the $\hbar\Omega=17$ MeV results in Fig. \ref{fig:hw17width} are clearly not converged yet. This suggests that the case of $\hbar\Omega=17$ MeV requires much larger model spaces to account for the physics that is already present at $\hbar\Omega=13$ MeV and $15$ MeV for the same $\Nmax$.  For these reasons, the $\hbar\Omega=17$ MeV case is excluded from the $\Nmax=9$ analysis and from the extrapolation procedure described below. Note that for the same  $\Nmax$, smaller \ho~values imply smaller ultraviolet (UV) cutoff (that resolves the high-momentum components of the interaction) and larger infrared  (IR) cutoffs (the size of the coordinate space in which the nucleus resides) \cite{Wendt_PRC91_2015}. Hence, the convergent results at $\hbar\Omega=13$ MeV suggest that to account for the physics of the resonance, spatially extended model spaces are imperative, which tracks with the nature of a cluster system.

In addition, ultra-large model spaces are accessible in the the NCSpM. The NCSpM results for $\Nmax=23$ (Fig. \ref{fig:hw15width}) may serve as a guidance for the convergence of the $\abinitio$ SA-NCSM results for $\hbar\Omega=15$ MeV. We note that the NCSpM is an effective approach and the value of $\hbar\Omega$ is fixed based on self-consistent arguments, which, in turn, has been shown to yield a close agreement to experimental observables, such as energies, quadrupole moments, $E2$ transitions, and radii in \nuc{20}{Ne} \cite{Tobin_PRC89_2014}. The present results reveal another remarkable outcome, namely,  the maximum of the NCSpM width coincides with the extrapolated SA-NCSM width (Fig. \ref{fig:hw13width}), as discussed next. 

To report a parameter-independent width, we use the Shanks transformation \cite{Schmidt_PM32_1941,Shanks_JMNP35_1955} on the $\Nmax=5,7,$ and 9 data for both the $\hbar\Omega=13$ MeV and $\hbar\Omega=15$ MeV results that are on a converging trend. With increasing $\Nmax$, the fastest convergence is observed for $\hbar\Omega=13$ MeV, for which the width flattens around $r_{c}=4.6$-$5$ fm (the fastest convergence indicates that there is an optimal \ho~value, where the convergence of results -- and the associated extrapolated estimate --  is achieved at comparatively lower  $\Nmax$, whereas lower or higher \ho~values require larger model spaces to yield the same estimate). Hence, this is the region where we perform extrapolation for both values of $\hbar\Omega$. While the extrapolation for $\hbar\Omega=13$ MeV is essentially independent of the channel radius (see black data in Fig. \ref{fig:widthsextrap}), the one for $\hbar\Omega=15$ MeV becomes unstable for larger $r_c$, indicating that larger model spaces are needed to attain the level of convergence necessary for the extrapolation. In such cases, due to the missing correlations in the model spaces considered, the extrapolation may yield lower estimates, as in Fig. \ref{fig:hw15width}. Using the $r_c$-independent part of these extrapolations, we report a value of $\Gamma_{\alpha}=10(3)$ eV for the alpha partial width of the $1^{-}$ resonance, with uncertainty given by the variation in $\hbar\Omega$. Given that no parameters are fitted to nuclear data in this study, this estimate agrees reasonably well with the $\Gamma_{\alpha}=28(3)$ eV width determined from experiment  \cite{PhysRevC.22.356}. We  note the importance of collective correlations, as evident in Fig. \ref{fig:hw13width} for  the $r_c$-independent region, namely, there is a  large increase in the alpha width as one goes from  the $\Nmax=3$ model space with surpassed correlations to  $\Nmax = 9$. 

The calculated alpha partial width
is associated with a unitless reduced width of $\theta^{2}_{1}=0.61(6)$. This compares reasonably well with previous GCM estimates of $\theta^{2}_{1}>0.54$ \cite{Horiuchi_PTP49_1973} 
and 0.54 at a channel radius of 5 fm \cite{KanadaEnyo_PTEP2014_2014b}, although a number of previous MCM results report a smaller reduced width $\theta^{2}_{1}<0.45$ \cite{Fujiwara_PTPSupp68_1980}.

For the ground state of \nuc{20}{Ne}, the ANC (\ref{eq:ANC}) characterizes the overall scale of the long-range $A$-particle wave function in the $l=0$ $\nuc{16}{O}+\alpha$ channel. The \nuc{20}{Ne} \gs~is calculated in the $\abinitio$ SA-NCSM [with \SU{3} basis] with different values of \ho. The ANC shows similar dependence on the parameter $\hbar\Omega$, as in the case of the alpha partial width, although the ANC results for all three $\hbar\Omega$ values are clearly either converged or nearly converged with respect to $\Nmax$  (Fig. \ref{fig:ANC}). As the $\Nmax=8$ and 10 data are nearly indistinguishable, we use the Shanks transformation on the $\Nmax=4,6,$ and 8 data for $\hbar\Omega=13$ MeV and  $\hbar\Omega=15$ MeV and for $r_{c}=4.3$\,-\,$4.8$ fm, in order to report a parameter-independent value for the ANC, as well as to provide confirmation of convergence with \Nmax. The range of channel radii considered provides a channel-independent region for the fastest convergence of the ANC with $\Nmax$ (Fig. \ref{fig:hw13anc}). This yields an extrapolated ANC of $C_{0}=3.4\pm1.2\times10^{3}\,\mathrm{fm}^{-1/2}$ for the \nuc{20}{Ne} \gs~in the $l=0$ $\nuc{16}{O}+\alpha$ channel, with uncertainty given by the variation in $\hbar\Omega$.

Similar to the $1^-$ width, it is interesting to note the agreement between the NCSpM and SA-NCSM results for $C_0$. 
In particular, the  $\Nmax=22$ NCSpM results for  $\ho=15$ MeV  yields an ANC of $C_{0}=3.3(1)\times10^{3}\,\mathrm{fm}^{-1/2}$, with uncertainty given by the $\sim6\%$ variation in $r_{c}$ (Fig. \ref{fig:hw15anc}). In the $\Nmax=14$ NCSpM, we also compute the ANC for the excited 4.25-MeV $4^{+}$ state in \nuc{20}{Ne}
to be $C_{4}=34(1)\times10^{3}\,\mathrm{fm}^{-1/2}$.
We find $C_{4}$ to be much larger than $C_{0}$, but we note that the $4^+$ state is only 0.48 MeV below the $\alpha+\nuc{16}{O}$ threshold.

\subsection{Reaction rate and XRB abundances}
\begin{figure}[h!]
    \centering
    \includegraphics[width=0.45\textwidth]{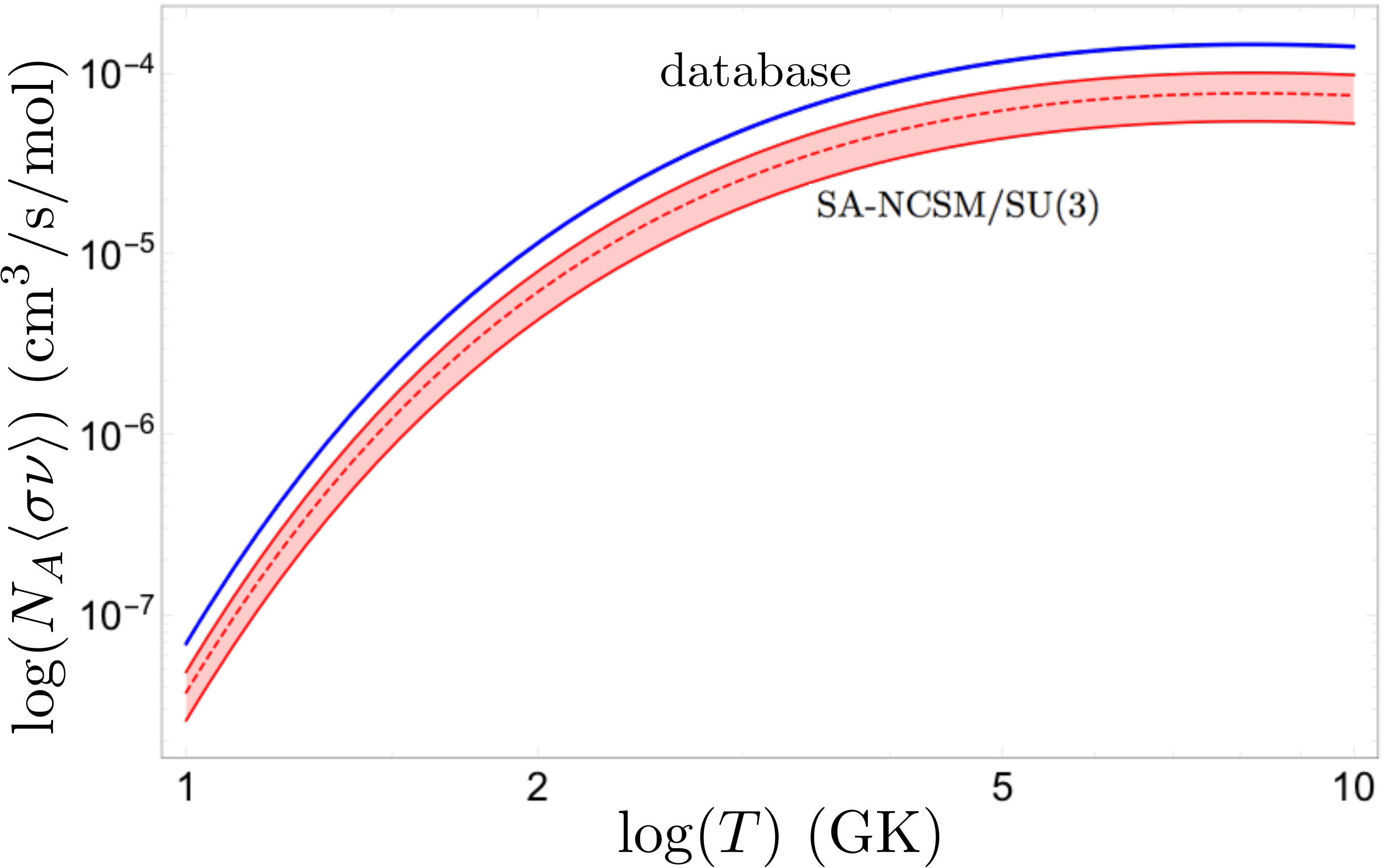}
    \caption
    {
        \label{fig:rxnrate} 
        Log-log plot of the $\nuc{16}{O}(\alpha,\gamma)\nuc{20}{Ne}$ reaction rate through the 1.06-MeV $1^{-}$ resonance in \nuc{20}{Ne} $(\mathrm{cm}^{3}/\mathrm{s}/\mathrm{mol})$ determined with Eq. (\ref{eq:narrowres}) as a function of the temperature (GK). The reaction rate determined with the extrapolated alpha partial width $\Gamma_{\alpha}=10(3)$ eV derived from $\abinitio$ SA-NCSM wave functions (red, labeled as ``SA-NCSM /\SU{3}") is compared to the database reaction rate (blue). The error in the database rate is given by the thickness in the curve.
    }
\end{figure}

Using the narrow resonance approximation, 
reaction rates are given by
\begin{equation}
    \label{eq:narrowres}
    N_{A}\langle\sigma\nu\rangle_{r}
    =
    \frac{1.539\times10^{11}}{(\mu_{A-a,a} T_{9})^{3/2}}
    e^{-11.605E_{r}/T_{9}}
    (\omega\gamma)_{r},
\end{equation}
with the resonance strength defined as
\begin{equation}
    (\omega\gamma)_{r}
    =
    \frac{2J_{r}+1}{(2J_{A-\alpha}+1)(2J_{\alpha}+1)}
    \frac{\Gamma_{\alpha}\Gamma_{\gamma}}{\Gamma}.
\end{equation}
We compute the temperature-dependent ($T_{9}$ in GK) contribution to the $\nuc{16}{O}(\alpha,\gamma)\nuc{20}{Ne}$ reaction rate through the 1.06-MeV $1^{-}$ resonance in \nuc{20}{Ne} 
(Fig. \ref{fig:rxnrate}).
The reaction rate takes as input the reduced mass $\mu_{A-a,a}$ (\ref{redmass}), resonance strength $(\omega\gamma)_{r}$ and resonance energy $E_{r}$ in MeV. Note that the resonance strength $(\omega\gamma)_{r}$ is dependent on the spins of the two clusters, $J_{\alpha}=0$ and $J_{^{16}{\rm O}}=0$ (or $J_{A-\alpha}$), as well as the spin of the narrow resonance, $J_{r}=1$. In addition, the resonance strength requires the alpha partial width $\Gamma_{\alpha}$, which we compute here, and the gamma decay branching ratio $\Gamma_{\gamma}/\Gamma$. This branching ratio is presently extracted from experiment \cite{Constantini_PRC82_2010}, namely, we adopt $\Gamma_{\gamma}/\Gamma=1.9\times10^{-4}$, but can be determined within this framework through SA-NCSM (or NCSpM) electromagnetic strengths. Because the branching ratio $\Gamma_{\gamma}/\Gamma$ and the resonance energy $E_{r}$ are kept constant for the two calculations, the differences in the reaction rates shown in Fig. \ref{fig:rxnrate} reflect the differences between the experimental and calculated alpha partial widths. 
\begin{figure*}[th]
    \includegraphics[width=0.8\textwidth]{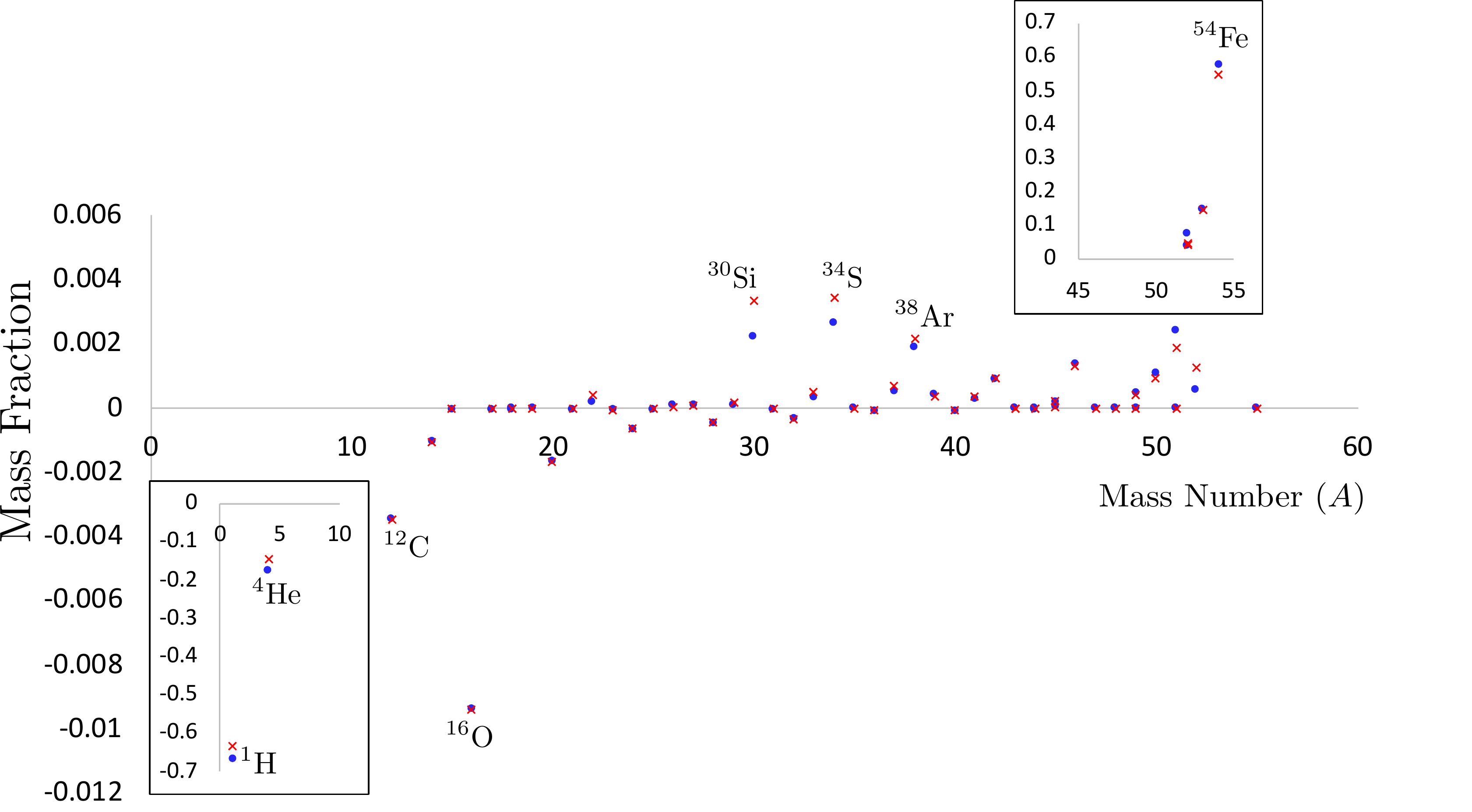}
    \caption
    {
        \label{fig:XRBabund}
        The difference between the initial mass fractions of the neutron star and the mass fractions 24 hours after the burst begins, based on the MESA XRB simulation that uses the database reaction rate (blue circles) and the reaction rate derived from $\abinitio$ SA-NCSM wave functions, shown in Fig. \ref{fig:rxnrate} (red $\times$). All isotopes in the network with mass differences greater than $10^{-10}$ are shown, and we label some isotopes of interest. The inset shows a detailed look at the abundance pattern for isotopes of H, He, Mn, and Fe. 
    }
\end{figure*}

Using this reaction rate as input to the Modules for Experiments in Stellar Astrophysics (MESA) code suite \cite{Paxton2011,Paxton2013,Paxton2015,Paxton2018,Paxton2019}, we are able to determine the impact on the abundance pattern produced during an X-ray burst (XRB) event using reaction data determined from $\abinitio$ wave functions (Fig. \ref{fig:XRBabund}). The MESA release \cite{Paxton2015} includes a model for an XRB with a constant accretion rate and consistent burning across the entire surface of the neutron star, based on GS 1826-24 \cite{Tanaka_ESASP296_1989}, also known as the ``clocked burster'' \cite{Ubertini_APJ514_1999}. This model is designed for a nuclear network of 305 isotopes, including proton rich isotopes up to \nuc{107}{Te}, but is also stable for a nuclear network of 153 isotopes. We use the 153-isotope nuclear network, which includes isotopes up to \nuc{56}{Fe}. MESA includes all known reactions involving these nuclei, with reaction data taken from the Joint Institute for Nuclear Astrophysics (JINA) REACLIB database \cite{Cyburt_APJS189_2010}. 

We examine the mass fractional abundances (i.e., masses given as a fraction of the total mass of the star) for a 24-hour period after the burst begins, for the theoretical rate (Fig. \ref{fig:rxnrate}), as compared to the database reaction rate. In this time frame, the system undergoes a number of bursts, but is sampled in a quiescent phase after a burst. The overall abundance pattern is relatively unchanged, except for a slight decrease in burning of the \nuc{1}{H} and \nuc{4}{He} fuels, which translates directly to a decrease in production of iron and manganese isotopes. This does not appear to be related to an overall change in the production of \nuc{16}{O} or \nuc{20}{Ne}, but does appear to slightly change the abundances of some other intermediate-mass nuclei, particularly \nuc{30}{Si}, \nuc{34}{S}, and \nuc{38}{Ar}, all of which have increased production with the change in the $\nuc{16}{O}(\alpha,\gamma)\nuc{20}{Ne}$ rate. Because of the slight reduction in the calculated reaction rate, alpha particles are apparently not burning as efficiently. As a result, some intermediate-mass nuclei are more abundant, while the production of Fe and Mn is reduced, but the overall pattern is only slightly affected. This means that theoretical predications  that use \abinitio wave functions are now possible and could provide reasonable estimates  for astrophysically relevant reaction rates that cannot be measured. 

%%%%%%%%%%%%%%%%%%%%
%%%% Conclusion %%%%
%%%%%%%%%%%%%%%%%%%%
\section{Conclusion}
In summary, we have outlined a new formalism for 
estimating
 alpha partial widths and ANCs from $\abinitio$ wave functions. We demonstrate  the formalism in a study of the $\nuc{16}{O}+\alpha$ cluster structure of \nuc{20}{Ne}, 
through inspection of the relative motion wave functions of the clusters within
the \nuc{20}{Ne} ground state and %more so for
 the $1_{1}^{-}$ resonance.   
 For the first time within a no-core shell model framework, we determine the alpha partial width for the 1.06-MeV $1^{-}$ resonance of $\alpha+\nuc{16}{O}$ from \abinitio SA-NCSM calculations of the \nuc{20}{Ne} states, and show it is in good agreement with
experiment.  We predict the ANC for the \nuc{20}{Ne} ground state as well as for the 4.25-MeV $4_{1}^{+}$ state. We highlight the importance of correlations in developing cluster structures.

Although the results presented here focus only on the $\nuc{16}{O}(\alpha,\gamma)\nuc{20}{Ne}$ alpha-capture reaction, the theory is not limited to spherical clusters as $^{16}$O and to
alpha clusters. Indeed, the theoretical framework is fully applicable to single-particle clusters (e.g., for studies of radiative proton capture), deuterons, and heavier clusters (the $\nuc{12}{C}+\nuc{12}{C}$ cluster system). The formalism is readily extensible to a number of generalizations, such as (1)  including vibrations in addition to the equilibrium shape to describe each of the clusters, (2) introducing more than one symplectic irrep, or several equilibrium shapes and their rotations and vibrations, to describe the composite system,  and (3) including multiple channels with different orbital momenta and spins of the clusters.

To illustrate the direct value of these calculations in astrophysics simulations, the calculated partial width is used in 
an initial
exploration of XRB nucleosynthesis. Because the $\nuc{16}{O}(\alpha,\gamma)\nuc{20}{Ne}$ reaction rate is dominated by the contribution through the $1^{-}$ resonance at XRB temperatures, the calculated width is used to characterize this contribution. In a MESA simulation of XRB nucleosynthesis, we find almost no difference on nuclear abundances as compared to the XRB simulation results when the experimental rate is used. This means that the present method, starting with \abinitio wave functions and without any parameter adjustments, enables reasonable predictions  for astrophysically relevant reaction rates that cannot be measured. In some cases, these estimates may represent a large improvement over existing database entries.

%%%%%%%%%%%%%%%%%%%%%%
\section*{Acknowledgments}
We thank A. Lauer for guiding us in the use of MESA, and for providing access to the XRB simulation we use. We also thank R. deBoer for valuable comments on the manuscript. This work benefitted from computing resources provided by Blue Waters, NERSC,  and the Center for Computation \& Technology at Louisiana State University. This work benefited from support by the National Science Foundation under Grant No. PHY-1430152 (JINA Center for the Evolution of the Elements). Support from the U.S. National Science Foundation (ACI -1713690, OIA-1738287, PHY-1913728), the Czech Science Foundation (16-16772S) and the Southeastern Universities Research Association are all gratefully acknowledged.  Part of this work was performed under the auspices of the DOE by Lawrence Livermore National Laboratory under Contract No. DE-AC52-07NA27344, with support from the U.S. Department of Energy, Office of Science, Office of Workforce Development for Teachers and Scientists, Office of Science Graduate Student Research (SCGSR) program, and from LLNL's LDRD program (16-ERD-022 and 19-ERD-017). The SCGSR program is administered by the Oak Ridge Institute for Science and Education (ORISE) for the DOE. ORISE is managed by ORAU under contract number DE-SC0014664.
%%%%%%%%%%%%%%%%%%%%%%

%%%%%%%%%%%%%%%%%%%%
%%%% APPENDICES %%%%
%%%%%%%%%%%%%%%%%%%%
\appendix

\section{Symplectic Lowering Operator}
\label{appB}

The symplectic lowering operator $\mathcal{B}^{(0\,2)}_{ \Lambda_{B}}$can be written in terms of the dimensionless HO raising and lowering operators, $b^{\dagger(1\,0)}_{xj}=\frac{1}{\sqrt{2}}(r_{xj}-\mathrm{ i} p_{xj})$  and $b^{(0\,1)}_{xj}$, respectively (similarly for $y$ and $z$), for each particle $j$ in an $A$-particle system. The position and momentum coordinates of the $j^{th}$ particle in the laboratory frame are $\vec{r}_{j}$ and $\vec{p}_{j}$, respectively (see Fig. \ref{fig:Coords}).

The lowering operator is written for the relative coordinates $\vec{\boldsymbol{\zeta}}$ (see, e.g., Refs. \cite{Tobin_PRC89_2014, Launey_PPNP89_2016})
\begin{align}
    \label{eq:raisingIMPORTANT}
    \mathcal{B}_{ \Lambda_{B}}^{(0\,2)}(\vec{\boldsymbol{\zeta}})
    & =
    \frac{1}{\sqrt{2}}
    \sum_{j=1}^{A}
    \{ b_{j}  \times b_{j}  \}^{(0\,2)}_{ \Lambda_{B}}
    -
    \frac{1}{\sqrt{2}A}
    \sum_{s,t=1}^{A}
    \{ b_{s}  \times b_{t}  \}^{(0\,2)}_{ \Lambda_{B}}
    \nonumber
    \\
    & =
    \mathcal{B}_{ \Lambda_{B}}^{(0\,2)}(\vec{\boldsymbol{r}})
    - 
    \mathcal{B}_{ \Lambda_{B}}^{(0\,2)}(\vec{R}).
\end{align}
It is obvious that the first sum can be divided into two sums, based on the particles in each cluster, and hence, $
\mathcal{B}_{ \Lambda_{B}}^{(0\,2)}(\vec{\boldsymbol{r}})
=
\mathcal{B}_{ \Lambda_{B}}^{(0\,2)}(\vec{\boldsymbol{r}}')
+
\mathcal{B}_{ \Lambda_{B}}^{(0\,2)}(\vec{\boldsymbol{r}}'')$
.

We intend to show that $\mathcal{B}^{(0\,2)}_{ \Lambda_{B}}$ 
can be written as $\mathcal{B}^{(0\,2)}_{ \Lambda_{B}}=\mathcal{B}^{(0\,2)}_{ \Lambda_{B},{\rm c}}+\mathcal{B}^{(0\,2)}_{ \Lambda_{B},{\rm rel}}$, or,
\begin{align}
    \centering
    \label{eq:endresult}
    \mathcal{B}^{(0\,2)}_{ \Lambda_{B}}(\vec{\boldsymbol{\zeta}})
    =
    \mathcal{B}^{(0\,2)}_{ \Lambda_{B}}(\vec{\boldsymbol{\zeta}'})
    +
    \mathcal{B}^{(0\,2)}_{ \Lambda_{B}}(\vec{\boldsymbol{\zeta}''})
    +
    \mathcal{B}^{(0\,2)}_{ \Lambda_{B}}(\vec{r}_{A-a,a}),
\end{align}
that is, a term that acts on the two clusters, and a second term that acts only on the relative motion between the two.

The coordinates $\vec{\boldsymbol{r}}'=\{\vec{r}_{1},\dots,\vec{r}_{A-a}\}$ and $\vec{\boldsymbol{r}}''=\{\vec{r}_{A-a+1},\dots,\vec{r}_{A}\}$ are the laboratory frame coordinates for the particles in each of the two clusters. Applying Eq. (\ref{eq:raisingIMPORTANT}) to each individual cluster, the $A$-particle lowering operator in Eq. (\ref{eq:raisingIMPORTANT}) becomes
\begin{align}
    \label{eq:secondline}
    \mathcal{B}_{ \Lambda_{B}}^{(0\,2)}(\vec{\boldsymbol{\zeta}})
    &=
    \mathcal{B}_{ \Lambda_{B}}^{(0\,2)}(\vec{\boldsymbol{\zeta}'})
    +
    \mathcal{B}_{ \Lambda_{B}}^{(0\,2)}(\vec{\boldsymbol{\zeta}''})
    \nonumber 
    \\
    &
    +
    \mathcal{B}_{ \Lambda_{B}}^{(0\,2)}(\vec{R'})
    +
    \mathcal{B}_{ \Lambda_{B}}^{(0\,2)}(\vec{R''})
    -
    \mathcal{B}_{ \Lambda_{B}}^{(0\,2)}(\vec{R}).
\end{align}
With $\mathcal{B}_{ \Lambda_{B}}^{(0\,2)}(\vec{\boldsymbol{\zeta}'})$ and $\mathcal{B}_{ \Lambda_{B}}^{(0\,2)}(\vec{\boldsymbol{\zeta}''})$
that act on the clusters, and using
\begin{equation}
    \mathcal{B}_{ \Lambda_{B}}^{(0\,2)}(\vec{r}_{A-a,a})=
        \mathcal{B}_{ \Lambda_{B}}^{(0\,2)}(\vec{R'})
        +
        \mathcal{B}_{ \Lambda_{B}}^{(0\,2)}(\vec{R''})
    -   
     \mathcal{B}_{ \Lambda_{B}}^{(2\,0)}(\vec{R}),
\end{equation}
which utilizes Eq. (\ref{eq:raisingIMPORTANT}) for two effective particles with laboratory frame coordinates $\vec{R'}$ and $\vec{R''}$, we obtain Eq. (\ref{eq:endresult}) for the lowering operator in relative coordinates for the $A$-particle system.

%======================
\section{Overlap of cluster and symplectic states}
\label{appA}
The spectroscopic amplitude in Eq. (\ref{eq:specamp}) is dependent on the overlap,
\begin{equation}
    \mathcal{O}
    =
   \langle
        \sigma {n} \rho \omega \Lambda 
        |
        (\omega_{\rm c}; \eta(\eta\,0))\rho \omega \Lambda
            \rangle,
\end{equation}
(with $\Lambda = \kappa LM_L$) between a symplectic wave function for the $A$-particle system and a cluster  wave function for the two-cluster system comprised of an $a$-particle cluster and an $(A-a)$-particle cluster (we omitted the dependence on the additional quantum numbers $\alpha$ and $\alpha_{\rm c}$, since the derivation is independent of them). Our aim is to determine a recursive expression for the overlap, so that, e.g., the term with $\eta$ HO total excitations in the relative motion is determined directly from the overlap for $\eta-2$ excitations, as prescribed in Ref. \cite{Suzuki_NPA448_1986}. To achieve this, we use the symplectic lowering operator $\mathcal{B}_{ \Lambda_{B}}^{(0\,2)}(\vec{\boldsymbol{\zeta}})$, which lowers a symplectic state by two HO excitations, acting on the cluster state:
\begin{align}
    &\mathcal{B}_{ \Lambda_{B}}^{(0\,2)}
    |
    (\omega_{\rm c0}; \eta_{0}(\eta_{0}\,0))\rho_{0}\omega_{0}\Lambda_{0} 
    \rangle
    = 
    \sum_{\omega_{\rm c}\omega_{\eta}\rho \omega \Lambda }
    |
    (\omega_{\rm c};\omega_{\eta})\rho \omega\Lambda 
    \rangle
    \nonumber
    \\
    &
    \langle
    (\omega_{\rm c};\omega_{\eta})\rho \omega\Lambda 
    |
    \mathcal{B}_{ \Lambda_{B}}^{(0\,2)}
    |
    (\omega_{\rm c0}; \eta_{0}(\eta_{0}\,0))\rho_{0} \omega_{0}  \Lambda_{0} 
    \rangle,
\end{align}
where we have introduced the completeness relation for the cluster basis on the right hand side. Projecting this onto a symplectic  state and inserting the symplectic basis completeness on the left hand side, we obtain the desired overlap relation between states with $N_{\omega_0}$ and $N_{\omega}=N_{\omega_0}-2$:
\begin{widetext}
    \begin{align}
        \label{eq:MainPfEq}
        &\sum_{n_{0}}
        \langle 
         \sigma  {n}  {\rho}  {\omega}  { \Lambda}
        |
        \mathcal{B}_{ \Lambda_{B}}^{(0\,2)}
        |
        \sigma n_{0}\rho_{0}\omega_{0} \Lambda_{0}
        \rangle
        \langle
        \sigma n_{0}\rho_{0}\omega_{0} \Lambda_{0}
        |
        (\omega_{\rm c0}; \eta_{0}(\eta_{0}\,0))\rho_{0} \omega_{0}  \Lambda_{0} 
        \rangle
        \nonumber
        \\
        &
        = 
        \sum_{\omega_{\rm c}\omega_{\eta}}
        \langle
        (\omega_{\rm c};\omega_{\eta})\rho \omega \Lambda 
        |
        \mathcal{B}_{ \Lambda_{B}}^{(0\,2)}
        |
        (\omega_{\rm c0}; \eta_{0}(\eta_{0}\,0))\rho_{0} \omega_{0}  \Lambda_{0} 
        \rangle
        \langle 
       \sigma  {n}  {\rho}  {\omega}  { \Lambda}
        |
        (\omega_{\rm c};\omega_{\eta})\rho \omega  \Lambda \rangle,
    \end{align}
\end{widetext}
where we have used that the overlap is nonzero only when the $\{\rho \omega \Lambda\}$ labels on both sides of the overlap are equal. Hence, we have an expression that relates the matrix element of the symplectic lowering operator in the symplectic basis to the matrix element of the same operator in the cluster basis. To use the overlap relation, these matrix elements need to be derived, as discussed next.

In order to determine the matrix element of the symplectic lowering operator in a cluster  state, we need to consider the coordinates in which the lowering operator is written. In the relative coordinates used by the NCSpM, the lowering operator can be separated into two pieces: a lowering operator that acts only on the clusters $\mathcal{B}_{ \Lambda_{B},\mathrm{c}}^{(0\,2)}$ and a lowering operator that acts only on the relative motion coordinate $\mathcal{B}_{ \Lambda_{B},\mathrm{rel}}^{(0\,2)}$ (see Appendix \ref{appB}). Using this, we can rewrite the matrix element of the lowering operator in the cluster basis into two terms
\begin{widetext}
    \begin{align}
        \label{eq:MEintwoterms}
        \langle
        (\omega_{\rm c};\omega_{\eta})\rho\omega\Lambda
        |
        \mathcal{B}_{ \Lambda_{B}}^{(0\,2)}
        &|
        (\omega_{\rm c0}; \eta_{0}(\eta_{0}\,0))\rho_{0}\omega_{0}\Lambda_{0} 
        \rangle
        =
        \sum_{\Lambda_{\rm c}\Lambda_{\eta}\Lambda_{\eta_{0}}\Lambda_{c_{0}}}
        \langle
        \omega_{\rm c} \Lambda_{\rm c} ; \omega_{\eta} \Lambda_{\eta} | \omega\Lambda
        \rangle
        \langle
        \omega_{\rm c0} \Lambda_{\rm c0} ; (\eta_{0}\,0) \Lambda_{\eta_{0}} | 
       \omega_{0} \Lambda_{0}
        \rangle
        \nonumber
        \\
        &
        \times
        [
            \langle
            \omega_{\rm c}\Lambda_{\rm c}
            | \mathcal{B}_{ \Lambda_{B},\,c}^{(0\,2)}
            |\omega_{\rm c0}\Lambda_{\rm c0}
            \rangle 
            \delta_{\omega_{\eta}, (\eta_{0}\,0)}
            \delta_{\Lambda_{\eta}\Lambda_{\eta_{0}}}
            +
            \langle
            \omega_{\eta} \Lambda_{\eta} 
            | \mathcal{B}_{ \Lambda_{B},\,\mathrm{rel}}^{(0\,2)} |
            \eta_{0}(\eta_{0}\,0)\Lambda_{\eta_{0}}
            \rangle
            \delta_{\omega_{\rm c}\omega_{\rm c0}}
            \delta_{\Lambda_{\rm c}\Lambda_{\rm c0}}
        ].
    \end{align}
\end{widetext}
For clusters 
with suppressed vibrations,
described by the bandhead of a symplectic irrep, $\mathcal{B}_{ \Lambda_{B},\,c}^{(0\,2)}|\omega_{\rm c} \Lambda_{\rm c}\rangle = 0$, and so only the second term is nonzero. While this is the case we consider here,  the present formalism can be generalized by using both terms in Eq. (\ref{eq:MEintwoterms}).  The second term represents the action of the symplectic lowering operator on the relative motion. After reducing the matrix element on the left hand side using the \SU{3} Wigner-Eckart Theorem \cite{Draayer_JMP14_1973}, and collecting reduced Wigner coefficients into an \SU{3} Racah coefficient $U$  \cite{Draayer_JMP14_1973}, Eq. (\ref{eq:MEintwoterms}) is expressed simply as 
\begin{align}
    \label{eq:MEclusterlowering}
    \langle
    (\omega_{\rm c};\omega_{\eta})\rho\omega
    &\| \mathcal{B}^{(0\,2)} \|
    (\omega_{\rm c_{0}}; \eta_{0}(\eta_{0}\,0))\rho_{0}\omega_{0}
    \rangle
    \nonumber
    \\
    &
    = 
    \langle
    \omega_{\eta}
    \| \mathcal{B}^{(0\,2)}\|
    \eta_{0}(\eta_{0}\,0)
    \rangle
    \delta_{\omega_{\rm c}\omega_{\rm c0}}
    \nonumber
    \\
    &
    \times
    U[\omega_{\rm c}(\eta_{0} 0)\omega(0\,2);\omega_{0}\rho_{0}1;\omega_{\eta}1\rho].
\end{align}

This is nonzero only when $\omega_{\eta}=\eta_0-2 (\eta_0-2 \,\, 0)$ and $\rho_{B}'=\rho_{B}=1$ (see Eq. (12) in Ref. \cite{Rowe_JPALettEd17_1984}), with \cite{Rosensteel_JMP21_1980, Suzuki_NPA448_1986}:
\begin{align}
    \langle\eta_{0}-2(\eta_{0}-2\,\,0)
    \| \mathcal{B}_{\mathrm{rel}}^{(0\,2)} \|
    \eta_{0}(\eta_{0}\,0)\rangle_{1}
    = \sqrt{\mathrm{dim}(\eta_{0}\,0)}.
\end{align}

Returning to Eq. (\ref{eq:MainPfEq}), the matrix element of the symplectic lowering operator between the symplectic states on the left hand side can be expressed as \cite{Rowe_PPNP37_1996}: 
\begin{widetext}
    \begin{align}
        \label{eq:MEsymplowering}
        \langle 
        \sigma n\rho\omega\Lambda
        &| \mathcal{B}_{ \Lambda_{B}}^{(0\,2)} |
        \sigma n_{0}\rho_{0}\omega_{0}\Lambda_{0}
        \rangle
        \nonumber
        \\
        &
        =
        (-)^{\omega - \omega_{0}}
        U[\sigma n_{0}\omega(0\,2);\omega_{0}\rho_{0} 1; n 1 \rho]
        \langle
        \omega_{0}\Lambda_{0} ; (0\,2)\Lambda_{B} | \omega\Lambda
        \rangle
        [\Delta \Omega_{K}(n_{0}\omega_{0}, n\omega)]^{1/2}
        ( n \| \mathcal{B}^{(0\,2)} \| n_{0} ),
    \end{align}
\end{widetext}
where the non-normalized reduced matrix element $( n \| \mathcal{B}^{(0\,2)} \| n_{0} )$ is computed using the expressions in Table I of Ref. \cite{Rosensteel_PRC42_1990}, and the normalization is introduced through the diagonal part of the $\mathcal{K}$-matrix \cite{Rowe_JPALettEd17_1984,Hecht_JPA18_1985}, as outlined in Eqs. (\ref{kmat1}) and (\ref{kmat2}).
Substituting Eqs. (\ref{eq:MEsymplowering}) and (\ref{eq:MEclusterlowering}) back into Eq. (\ref{eq:MainPfEq}), we obtain the relation in Eq. (\ref{eq:recursion}) for determining the overlap between cluster and symplectic states.

%%%%%%%%%%%%%%%%%%%%
%%% BIBLIOGRAPHY %%%
%%%%%%%%%%%%%%%%%%%%
\bibliographystyle{apsrev}
\bibliography{Ne20.bib}

\end{document}